\documentclass[aps,reprint,groupedaddress,showpacs,onecolumn,notitlepage]{revtex4-1}

\usepackage{amsmath,amssymb}
\usepackage{graphicx}
\usepackage{array}
\usepackage{bm}
\usepackage{subfigure}

\begin{document}

\title{Stability of arrays of bottom-heavy spherical squirmers}
\author{D. R. Brumley$^{1}$ and T. J. Pedley$^{2}$}

\email{d.brumley@unimelb.edu.au, T.J.Pedley@damtp.cam.ac.uk}

\affiliation{$^1$School of Mathematics and Statistics, The University of Melbourne, Victoria 3010, Australia. \\
$^2$Department of Applied Mathematics and Theoretical Physics, Centre for 
Mathematical Sciences, University of Cambridge, Wilberforce Road, Cambridge CB3 0WA, 
UK}
\date{\today}

\begin{abstract}
The incessant swimming motion of microbes in dense suspensions can give rise to striking collective motions and coherent structures. However, theoretical investigations of these structures typically utilize either computationally demanding numerical simulations, or simplified continuum models. Here we analytically investigate the collective dynamics of a dense array of steady, spherical squirmers. We first calculate the forces and torques acting on two closely-separated squirmers, through solving the Stokes equations to second-order in the ratio of mean spacing to squirmer radius. This lubrication analysis is then used to assess the stability of a dense, vertical, planar array of identical three-dimensional squirmers. The system of uniformly-spaced, vertically-oriented squirmers is stable if the gravitational torque experienced due to bottom-heaviness is sufficiently strong, and an intercellular repulsive force is included. The predictions of instability and possible long time behavior is qualitatively similar for monolayers confined between two parallel rigid planes as for unconfined monolayers. The predictions compare favourably with published numerical simulations, and reveal the existence of additional dynamic structures not previously observed; puller-type squirmers show a greater range of structures than pushers. The use of pairwise lubrication interactions provides an efficient means of assessing stability of dense suspensions usually tackled using full numerical simulations.
\end{abstract}

\maketitle

\section*{Introduction}

Motility is a pervasive feature among microorganisms, from the diurnal migration of marine phytoplankton \cite{Bollens2010} to the motion of bacteria in the gut \cite{Berg2008}. The acquisition of resources \cite{Blackburn1998}, evasion from predators \cite{Kiorboe:2014aa}, and infection by pathogens \cite{Josenhans2002} all depend sensitively on organismal motility. Early microscopes dating back to the 18th century \cite{VLeeuwenhoek:1700} enabled glimpses into the dynamic nature of the microbial world. Since then, the role of cilia and flagella -- ubiquitous, highly conserved propulsive appendages -- has received considerable attention \cite{Sleigh1962, Brennen:1977}. Recent advances in imaging and microfluidic control offer new insights into the mechanics of cellular propulsion \cite{Son2015}. The spatial distribution of cells in microbial consortia can influence nutrient cycling \cite{Smriga2016}, horizontal gene transfer \cite{Moor2017} and fertilization processes \cite{Denissenko2012}. Developing a quantitative framework for the collective dynamics of swimming microorganisms is therefore essential to understanding a vast array of biological processes. It has become clear that collective motions of many microorganisms can be very different from individual dynamics \cite{Elgeti2015}. Striking examples of bacterial turbulence \cite{Dunkel:2013}, self-organization \cite{Wioland2013, Thutupalli2018} and coherent structures \cite{Saintillan:2012zr} reveal the combined effects of confinement, hydrodynamic signatures, and steric interactions in determining emergent phenomena.

The squirmer model was first proposed by Lighthill in 1952 \cite{Lighthill:1952} and modified by his student Blake in 1971 \cite{Blake:Squirmer}, but its current wide applicability to a range of organisms was not initiated until relatively recently \cite{Pedley2016b, Gilpin2017, Shapiro2014, Magar:2005, Michelin2011, Lin2011}. Its elegance and simplicity enable modelling of cells in different environments \cite{Matas-Navarro2014}, near air-liquid interfaces \cite{Wang2013}, or no-slip and repulsive walls \cite{Li2014, Llopis:2010, Lintuvuori2016}. The conceptually simple model replaces an array of flagella with a single, no-slip, deformable surface, thereby linking discrete ciliary beating with an effective surface slip velocity. The model organism, {\it Volvox carteri} \cite{Goldstein2015}, renowned for its exquisite spherical symmetry, exemplifies the squirmer model, with strong agreement between predictions based on measured flagellar dynamics \cite{Pedley2016, Brumley2014, Brumley2015} and the observed motion of freely-swimming colonies \cite{Drescher:2010kx}. Experimental \cite{Drescher:2009vn} and theoretical evidence \cite{Sano2016} hints at the importance of near-field interactions in determining collective properties of suspensions of squirmers.

Ishikawa {\it et al.} \cite{Ishikawa2006} investigated hydrodynamic interactions between two spherical squirmers, utilizing both lubrication theory and multipole expansions to model closely- and widely- separated squirmers respectively. Boundary element simulations of dense suspensions revealed stable collective states and intriguing oscillatory modes \cite{Ishikawa2008a, Ishikawa2008}, in which squirmers self organize into a densely packed lattice. Despite the conceptual simplicity of the squirmer model, it remains unclear what mechanisms are responsible for these states, and the precise conditions under which they are stable. In this paper, we analytically solve the Stokes equations between two bottom-heavy, spherical squirmers, in the limit of close separation. We use these results to predict the collective dynamics of a dense array of squirmers, and show that both orientational and translational stability are mediated through gravitational torques exerted on the cells, and a cell-cell repulsive force. 

\section{Interactions between spherical squirmers}

\subsubsection{Interactions due to squirming motion} \label{interaction_squirm}

In order to calculate the forces and torques arising from the short-range interactions between two spherical swimming microorganisms, we will utilize the squirmer model. The single-squirmer model will be developed in the reference frame in which the center of the spherical squirmer is at rest, and the fluid at infinity has velocity given by $-U \bm{e}$. The value $U$ is the swimming speed of the sphere and $\bm{e}$ is its orientation vector -- the unit vector along the axis of symmetry. The boundary conditions at the surface of the sphere are given by
\begin{equation}
u_r \big|_{r=a} = \sum_{n} A_n(t) P_n (\cos \theta), \qquad u_{\theta} \big|_{r=a} = \sin \theta \sum_{n} B_n(t) W_n (\cos \theta) \label{squirmer_BC},
\end{equation}
where $\theta$ is the angle measured from the anterior of the squirmer, $P_n$ is the $n^{\text{th}}$ Legendre polynomial, and $W_n$ is defined as
\begin{equation}
W_n(\cos \theta) = \frac{2}{n(n+1)} P_n' (\cos \theta).
\end{equation}

\begin{figure}[htp] \begin{center}
	\includegraphics[width=5cm]{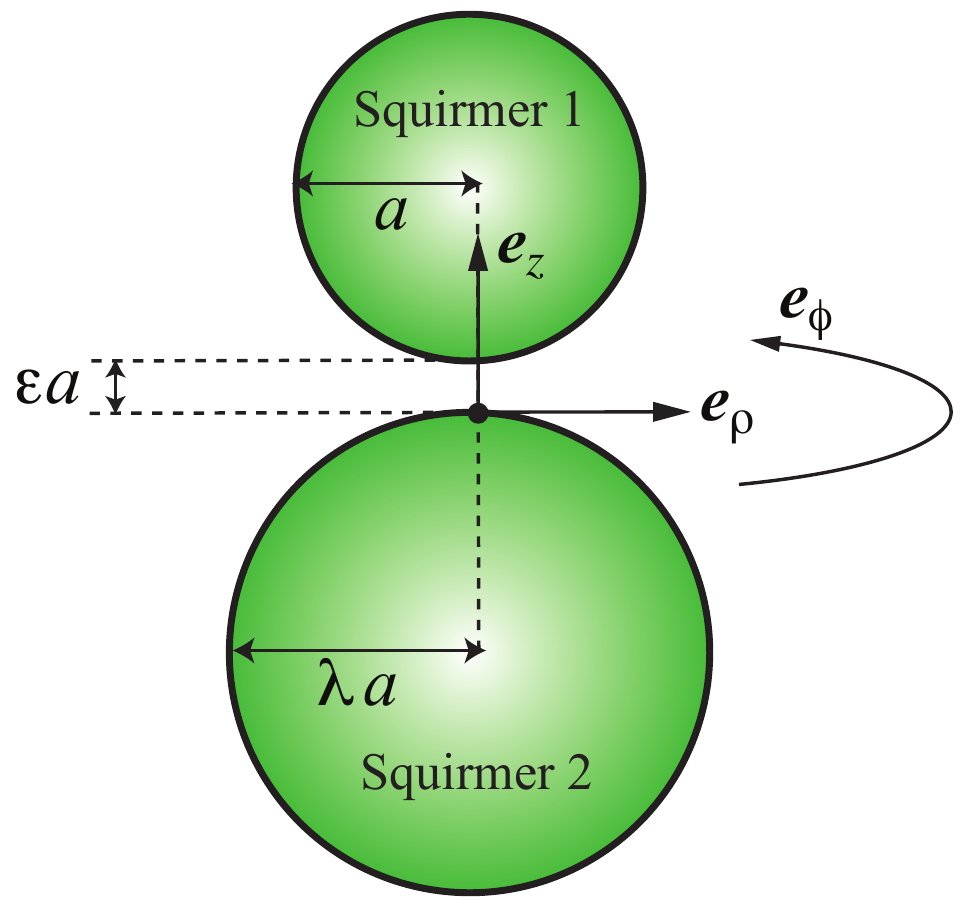}
	\caption{Geometry of the two squirmers. The origin of the coordinate system is located on the surface of sphere 2 closest to sphere 1. The vector $\bm{e}_{\rho}$ points radially in the $x$-$y$ plane, and the vector $\bm{e}_{\phi}$ is the azimuthal direction.}  \label{squirmer_pair}
\end{center} \end{figure}

The ultimate goal will be to consider the hydrodynamic interaction between two adjacent squirmers whose positions and orientations are arbitrary. Without loss of generality, consider the problem of two closely-separated spherical squirmers, as depicted in Fig.~\ref{squirmer_pair}. The frame is chosen such that the orientation of squirmer 1 lies in the $x$-$z$ plane. ie $\bm{e}_1 \cdot \bm{e}_y = 0$. Any configuration in a laboratory frame can be mapped to the situation shown in Fig.~\ref{squirmer_pair} through a suitable linear transformation. By linearity of the Stokes equations, the problem involving two squirming spheres in a fluid that is at rest infinitely far away can be broken down into two distinct problems. The first has the squirming-sphere boundary condition on sphere 1 and zero velocity boundary condition on sphere 2. The second problem has zero velocity on sphere 1 and the squirming-sphere boundary condition on sphere 2. Only the former problem will be studied, since solving this will immediately yield the solution to the latter. 

The radii of spheres 1 and 2 are given by $a$ and $\lambda a$ respectively, and the minimum separation between the spheres is taken to be $\epsilon a$ (with $\epsilon \ll 1$). The origin of the coordinate system is located at the surface of sphere 2, on the axis joining the centers of the two spheres. The $z$-axis passes through the spheres' centers, so that spheres 1 and 2 lie in the regions $z>0$ and $z \leq 0$ respectively. The surfaces of spheres 1 and 2 are determined by $z=h_1$ and $z=h_2$, respectively. Let the two spheres, 1 and 2, have orientation vectors $\bm{e}_1$ and $\bm{e}_2$ and squirming sets $\bm{B}^{(1)} = (B_1^{(1)}(t),B_2^{(1)}(t),\ldots)$ and $\bm{B}^{(2)} = (B_1^{(2)}(t),B_2^{(2)}(t),\ldots)$, respectively. Squirmers with zero radial velocity on the sphere surface will be considered ($A_n(t) = 0 \; \forall \; n$).
Although Fig.~\ref{squirmer_pair} depicts a configuration with two spheres, the following lubrication analysis can also be applied to the interaction between a sphere and a plane wall by considering the case where $\lambda \rightarrow \infty$.

\begin{figure}[h!]
\begin{center}
\includegraphics[width=0.75\textwidth]{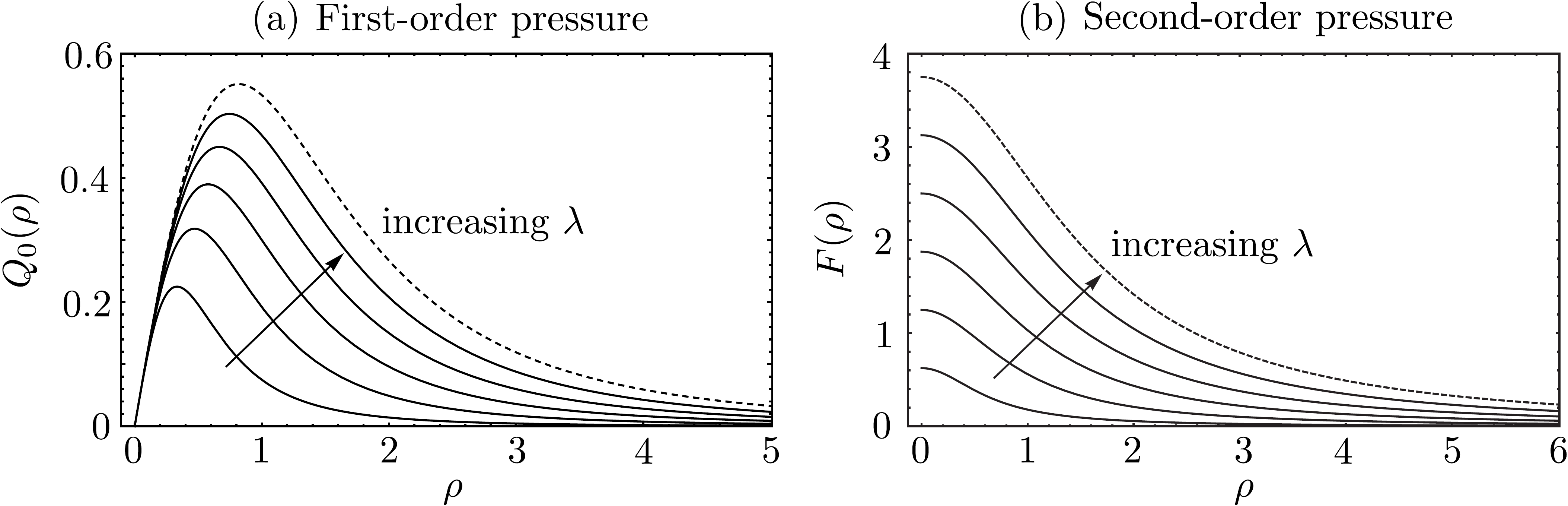}
\caption{First and second-order components of the pressure distribution in the lubrication region. (a) $Q_0(\rho)$ and (b) $F(\rho)$ are shown as functions of $\rho$. Results are shown for $\lambda=0.2$, 0.5, 1, 2, 5 and $\infty$ (dotted).} 
\label{pressure_first_and_second_order}
\end{center}
\end{figure}

The fluid velocity $\bm{u} = (u,v,w)$ and pressure $p$ in the gap between the squirmers, are expanded in powers of $\epsilon$:
\begin{align}
\begin{split}
u &= u_0 + \epsilon^{1/2} u_1 + \mathcal{O}(\epsilon), \\
v &= v_0 + \epsilon^{1/2} v_1 + \mathcal{O}(\epsilon), \\
w &= \epsilon^{1/2} w_0 + \epsilon w_1 +  \mathcal{O}(\epsilon^{3/2}), \\
p &= \epsilon^{-3/2} p_0 +  \epsilon^{-1} p_1 + \mathcal{O}(\epsilon^{-1/2}).
\end{split}
\label{velocity_expansions}
\end{align}
Similarly, the separation between the squirmers, $H$, non-dimensionalized by $\epsilon a$, can be written as a function of $\rho$ for $\epsilon \ll 1$,
\begin{equation}
H = 1 + \frac{\lambda+1}{2 \lambda} \rho^2 + \mathcal{O}(\epsilon),
\end{equation}
where $\rho$ is the distance from the $z$-axis (see Fig.~\ref{squirmer_pair}). By expanding and solving the Stokes equations in powers of $\epsilon$, the leading order pressure distribution $p_0 (\rho,\phi) = q_0 (\rho) \bm{e} \cdot \bm{e}_{\rho}$ can be found (see SI Section~S1
for detailed calculation), where
\begin{equation}
q_0 (\rho) = Q_0(\rho) \sum_n B_n W_n \big( -\bm{e} \cdot \bm{e}_z \big) \quad \text{and} \quad Q_0(\rho) = \frac{6 \mu}{5 a} \frac{\rho}{H^2}. \label{pressure_leading}
\end{equation}
Similarly, the second-order pressure is found to be of the form
\begin{equation}
p_1 (\rho,\phi) = f_{\text{p}}(\rho) + g(\rho) \cos 2\phi.
\end{equation}
The component proportional to $\cos 2\phi$ disappears upon integration with respect to $\phi$ and so does not provide a net contribution to the force exerted between the spheres. It therefore suffices to consider 
\begin{equation}
F(\rho) = \frac{3}{4} \bigg( \frac{\lambda}{\lambda+1} \bigg) \frac{6H-1}{H^2}, \label{pressure_second}
\end{equation}
where $F(\rho)$ is defined by
\begin{equation}
F(\rho) \sum_n \bigg[ B_n W_n \big( - \bm{e} \cdot \bm{e}_z \big) \bm{e} \cdot \bm{e}_z + \frac{1}{2} B_n W_n' \big( - \bm{e} \cdot \bm{e}_z \big) ( \bm{e} \cdot \bm{e}_{x} )^2 \bigg] = \frac{a}{\mu} f_{\text{p}}(\rho).
\end{equation}
The functions $Q_0(\rho)$ and $F(\rho)$ represent first- and second-order pressure increases due to the squirming motion of sphere 1. These are shown in Figs.~\ref{pressure_first_and_second_order}(a-b) respectively.

Using Eqs.~\eqref{pressure_leading} and \eqref{pressure_second}, the fluid velocity in the gap between the squirmers, Eq.~\eqref{velocity_expansions}, can be solved to first-order (see SI Section~S1).
These expressions enable the forces and torques acting on the two spheres to be calculated explicitly.
\begin{align}
\begin{split}
F_x^{(1)} &= -\frac{4}{5} \mu \pi a \ \bm{e} \cdot \bm{e}_x \frac{\lambda (\lambda+4)}{(\lambda+1)^2} \sum_n B_n W_n \big( -\bm{e} \cdot \bm{e}_z \big)  \big( \log \epsilon + \mathcal{O}(1) \big), \\
F_z^{(1)} &= -9 \mu \pi a \frac{\lambda^2}{(\lambda+1)^2} \sum_n \bigg[ B_n W_n \big( - \bm{e} \cdot \bm{e}_z \big) \bm{e} \cdot \bm{e}_z + \frac{1}{2} B_n W_n' \big( - \bm{e} \cdot \bm{e}_z \big) ( \bm{e} \cdot \bm{e}_{x} )^2 \bigg] \big( \log \epsilon + \mathcal{O}(1) \big), \\
T_y^{(1)} &= \frac{16 \lambda}{5 (\lambda+1)} \mu \pi a^2 \ \bm{e} \cdot \bm{e}_x \sum_n B_n W_n \big( -\bm{e} \cdot \bm{e}_z \big) \big( \log \epsilon + \mathcal{O}(1) \big), \\
T_y^{(2)} &= \frac{4 \lambda^2}{5 (\lambda+1)} \mu \pi a^2 \ \bm{e} \cdot \bm{e}_x \sum_n B_n W_n \big( -\bm{e} \cdot \bm{e}_z \big) \big( \log \epsilon + \mathcal{O}(1) \big).
\label{forces_and_torques}
\end{split}
\end{align}
The tangential and normal forces acting on sphere two, $F_x^{(2)} = - F_x^{(1)}$ and $F_z^{(2)} = - F_z^{(1)}$ respectively, are equal and opposite to the values on sphere one. By symmetry, the torque $T_x$ is precisely equal to zero. The torque in the $z$-direction can be evaluated; however it is found that $T_z = \mathcal{O}(\epsilon)$ so this need not be pursued. The torque exerted on sphere 2 in the $y$-direction has an extra factor of $\lambda$ compared to the results for sphere 1, arising from the discrepancy between their radii. It is also worth noting that for $\lambda = 1$ (equally-sized spheres), the torque exerted on sphere 2 is one quarter times that exerted on sphere 1. Normal gradients in the fluid velocity are greater at the surface of the squirmer than they are at the boundary of the no-slip sphere.

\subsubsection{Interactions due to motion of spheres} \label{interaction_motion}

In addition to the effects of squirming, the two spheres will also experience forces and torques due to their linear and angular velocities, intercellular steric interactions, and gravity. The analysis so far has been performed in the frame shown in Fig.~\ref{squirmer_pair}. The unit vectors appearing in Eq.~\eqref{forces_and_torques} will now be given the primed coordinates to indicate that they are viewed in this frame. Suppose that in the reference frame $S'$ depicted in Fig.~\ref{squirmer_pair}, the spheres possess linear and angular velocity vectors $\bm{V}_i' = (V_{x,i}',V_{y,i}',V_{z,i}')$ and $\bm{\omega}_i' = (\omega_{x,i}', \omega_{y,i}', \omega_{z,i}')$ respectively. The subscript $i=1,2$ denotes either sphere 1 or sphere 2. Let $\bm{F}_i'$ and $\bm{T}_i'$ be the force and torques acting on sphere $i=1,2$ in this frame $S'$. The forces and torques are scaled according to $\bar{\bm{F}}' = \bm{F}' /(\mu \pi a)$ and $\bar{\bm{T}}' = \bm{T}'/(\mu \pi a^2)$. The following relationships can then be established \cite{KimAndKarrila:Microhydrodynamics}:
\begin{align}
\bar{\bm{F}}_1' &= \textbf{A}  \cdot \big( \bm{V}_1' - \bm{V}_2' \big) +  \textbf{C}  \cdot \big( a \bm{\omega}_1' + a \bm{\omega}_2' \big),  \label{F1_s-s_veldep} \\
\bar{\bm{T}}_1' &= - \textbf{C} \cdot \big( \bm{V}_1' - \bm{V}_2' \big) +  \textbf{J} \cdot \bigg( \frac{8}{5} a \bm{\omega}_1' + \frac{2}{5} a \bm{\omega}_2'  \bigg),  \label{T1_s-s_veldep} \\
\bar{\bm{F}}_2' &= -\textbf{A} \cdot \big( \bm{V}_1' - \bm{V}_2' \big) - \textbf{C} \cdot \big( a \bm{\omega}_1' +a  \bm{\omega}_2' \big),   \label{F2_s-s_veldep} \\
\bar{\bm{T}}_2' &= - \textbf{C} \cdot \big( \bm{V}_1' - \bm{V}_2' \big) +  \textbf{J} \cdot \bigg( \frac{2}{5} a \bm{\omega}_1' + \frac{8}{5} a \bm{\omega}_2'  \bigg),  \label{T2_s-s_veldep}
\end{align}
where, correct to order $\mathcal{O}(\log \epsilon)$, the matrices are given by
\begin{equation}
\textbf{A} = \left( \begin{array}{c c c}
\log \epsilon & 0 & 0 \\
0 & \log \epsilon & 0 \\
0 & 0 & -\frac{3}{2 \epsilon} + \frac{27}{20} \log \epsilon \\
\end{array} \right), \qquad
\textbf{C} = \left( \begin{array}{c c c}
0 & -\log \epsilon & 0 \\
\log \epsilon & 0 & 0 \\
0 & 0 & 0 \\
\end{array} \right), \qquad
\textbf{J} = \left( \begin{array}{c c c}
\log \epsilon & 0 & 0 \\
0 & \log \epsilon & 0 \\
0 & 0 & 0 \\
\end{array} \right).
\end{equation}
As expected, $\bar{\bm{F}}_2' = - \bar{\bm{F}}_1'$ and the forces and torques arising due to linear velocities are zero when $\bm{V}_1' - \bm{V}_2'  = 0$. Note that these results correspond to the case involving two equally-sized spheres ($\lambda=1$). Rotation of the spheres in the $z$-direction does not produce forces or torques that are singular as $\epsilon \rightarrow 0$, and the torque in the $z$-direction remains finite as the spheres become arbitrarily close together. It follows that the entries in the third row of $\textbf{C}$ and $\textbf{J}$ are all zero to order $\mathcal{O}(\log \epsilon)$.

\subsubsection{Additional interactions} \label{interaction_additional}

If the squirmers are bottom-heavy, there is an additional external torque acting on each sphere due to gravity. For species such as {\it Volvox}, this mechanism facilitates swimming in an upwards direction (negative gravitaxis). If the distance between the center of gravity and center of the squirmer is given by $h$, in the direction opposite to its swimming direction in an undisturbed fluid, the gravitational torque on the $i^{\text{th}}$ squirmer is given by
\begin{equation}
\bm{T}_{\text{grav}}^i = - \frac{4}{3} \pi a^3 \rho_f h \ \bm{e}_i \times \bm{g},
\end{equation}
where $\rho_f$ is the density and $\bm{g}$ is the acceleration due to gravity. The parameter introduced by Ishikawa {\it et al.} \cite{Ishikawa2006} is adopted here, comparing the gravitational and viscous torques:
\begin{equation}
G_{bh} = \frac{2 \pi \rho_f g a h}{\mu B_1}. \label{G_bh_def}
\end{equation}
The non-dimensionalized gravitational torque can then be rewritten as $\bar{T}^i_{\text{grav}} = T^i_{\text{grav}}/(\mu \pi a^2) =  -\frac{2}{3 \pi}  B_1 G_{bh} \sin \zeta_i  \label{T_grav_i}$, where $\zeta_i$ is the angle of the squirmer from vertical.
A repulsive force between spheres was included in the numerical simulations of ref \cite{Ishikawa2006}:
\begin{equation}
\bm{F}_{\text{rep}} = \kappa_1 \kappa_2 \frac{\exp(-\kappa_2 \epsilon)}{1-\exp(-\kappa_2 \epsilon)} \frac{\bm{r}}{r}. \label{repulsive_force}
\end{equation}
This was done to avoid the prohibitively small time step required to prevent squirmers from overlapping. The parameter $\epsilon$ is again the separation between squirmers, non-dimensionalized by the squirmer radius $a$. The parameter $\kappa_1$ represents the strength of the repulsion while $\kappa_2$ dictates the range at which this repulsion becomes significant. The values adopted will be the same as those used previously \cite{Ishikawa2006}, namely $\kappa_1 = 1$ and $\kappa_2 = 10^3$. Importantly, this repulsive force can be ``switched off'' simply by choosing $\kappa_1 = 0$.

\subsubsection{Generalization to multiple spheres} \label{lubrication_multiple_spheres}

A larger system containing $n$ spheres will now be examined, and the total force and torque acting on each sphere will be calculated. Each sphere will interact hydrodynamically with neighbors that are sufficiently close, experience a gravitational torque due to bottom-heaviness, and be subject to the short-range repulsive force. The analysis so far has been performed in the frame shown in Fig.~\ref{squirmer_pair}. The unit vectors will now be given the primed coordinates to indicate that they are viewed in this frame. In accordance with the preceding lubrication theory, the squirmer is required to be oriented in the $x'$-$z'$ plane. That is, $\bm{e}_i \cdot \bm{e}_y' = 0$. For any ordered pair of spheres $i$ and $j$ in the laboratory frame $S$, it is possible to transform to the reference frame $S_{ij}'$ in which spheres $i$ and $j$ are positioned as squirmers 1 and 2 respectively in Fig.~\ref{squirmer_pair}. Suppose that in the laboratory reference frame $S$, two squirmers $i$ and $j$ have position vectors $\bm{r}_i$ and $\bm{r}_j$ respectively and that their orientations are given by $\bm{e}_i$ and $\bm{e}_j$ respectively. Let $\bm{r} = \bm{r}_i - \bm{r}_j$. The coordinate system $S_{ij}'$ is defined in the following way:
\begin{align}
\bm{e}_z' &= \hat{\bm{r}}, \\
\bm{e}_y' &= \hat{\bm{s}}, \quad \text{where} \quad \bm{s} = \bm{e}_z' \times \bm{e}_i, \\
\bm{e}_x' &= \bm{e}_y' \times \bm{e}_z'.
\end{align}
By construction, this frame satisfies the condition that $\bm{e}_i \cdot \bm{e}_y' = 0$. Thus, the lubrication analysis presented in Section~\ref{interaction_squirm} can be directly applied in this frame. In the calculation of the forces and torques due to squirming, it is also necessary to know the quantities
\begin{equation}
\bm{e}_i \cdot \bm{e}_z' = \bm{e}_i \cdot \hat{\bm{r}}, \qquad \text{and} \qquad \bm{e}_i \cdot \bm{e}_x' = \sqrt{1-(\bm{e}_i \cdot \hat{\bm{r}})^2}.
\end{equation}
The lubrication analysis is used in frame $S_{ij}'$ for the case when sphere $i$ is a squirmer and sphere $j$ has the zero boundary condition. The complementary problem involving sphere $j$ as a squirmer and sphere $i$ as a sphere with zero boundary condition is considered separately in frame $S_{ji}'$  since the lubrication analysis is only applicable when $\bm{e}_j \cdot \bm{e}_y' = 0$. In this fashion, each pair of squirmers will be considered twice when calculating the total force and torque on the system.

Equations~\eqref{F1_s-s_veldep}-\eqref{T2_s-s_veldep} outline the forces and torques due to translational and rotational velocities of the two spheres, where everything is measured in the frame $S_{ij}'$. However, it is desirable to find these quantities in the laboratory frame $S$. This is achieved by utilizing the appropriate transformation matrix, $\textbf{R}_{ij}$. A complete matrix-vector equation can be assembled as follows
\begin{align}
\left( \begin{array}{c}
\bar{\bm{F}}_1 \\
\vdots \\
\bar{\bm{F}}_n \\
\hline
\bar{\bm{T}}_1 \\
\vdots \\
\bar{\bm{T}}_n \\
\end{array} \right)^{\text{net}}
&=
\left( \begin{array}{c|c}
\textbf{M}_1 & \textbf{M}_2 \\
\hline
\textbf{M}_3 & \textbf{M}_4 \\
\end{array} \right)
\left( \begin{array}{c}
\bm{V}_1 \\
\vdots \\
\bm{V}_n \\
\hline
a \bm{\omega}_1 \\
\vdots \\
a \bm{\omega}_n \\
\end{array} \right)
+
\left( \begin{array}{c}
\bar{\bm{F}}_1 \\
\vdots \\
\bar{\bm{F}}_n \\
\hline
\bar{\bm{T}}_1 \\
\vdots \\
\bar{\bm{T}}_n \\
\end{array} \right)^{\text{sq}}
+
\left( \begin{array}{c}
\bar{\bm{F}}_1 \\
\vdots \\
\bar{\bm{F}}_n \\
\hline
0 \\
\vdots \\
0 \\
\end{array} \right)^{\text{rep}}
+
\left( \begin{array}{c}
0 \\
\vdots \\
0 \\
\hline
\bar{\bm{T}}_1 \\
\vdots \\
\bar{\bm{T}}_n \\
\end{array} \right)^{\text{grav}}. \label{all_forces_and_torques}
\end{align}
Since the fluid is considered to be at zero Reynolds number, the net force and torque on every squirmer must be zero. This condition is imposed simply by setting every entry on the left hand side of Eq.~\eqref{all_forces_and_torques} to zero. The resulting matrix-vector equation can then be solved to find the linear and angular velocities corresponding to this condition. In particular, a system of the following form must be solved:
\begin{equation}
\textbf{M} \cdot \bm{x} = \bm{R}, \label{matrix-vector_equation3}
\end{equation} 
where $\bm{x}$ contains the linear and angular velocities of the spheres. There are several important features of this equation that will now be discussed. Firstly, note that the matrix $\textbf{M}$ depends only on the positions of the squirmers. It is completely independent of the squirming parameters, the strength of gravity, the orientations of the squirmers and the repulsive force. Matrix $\textbf{M}$ can be assembled once the physical configuration of the suspension is known. The vector $\bm{R}$ in Eq.~\eqref{matrix-vector_equation3} depends on all of the parameters involved in the problem. Secondly, note that the $6n \times 6n$ matrix $\textbf{M}$ has a rank which is precisely $6n-3$. In order to understand this, recall that the only hydrodynamic forces and torques are those arising from the lubrication regions, which depend on the relative motion of the squirmers compared to each other. In order to ensure that the matrix $\textbf{M}$ is nonsingular, an arbitrary reference frame must be chosen. The frame in which the bulk velocity of the configuration of squirmers is zero is chosen. That is,
\begin{equation}
\sum_{i=1}^n V_{i,x} = \sum_{i=1}^n V_{i,y} = \sum_{i=1}^n V_{i,z} = 0. \label{net_velocity_zero}
\end{equation}

\section{Uniform Monolayer of squirmers in an unbounded fluid}

\begin{figure}[htp]
  \begin{center}
    \includegraphics[width=0.75\textwidth]{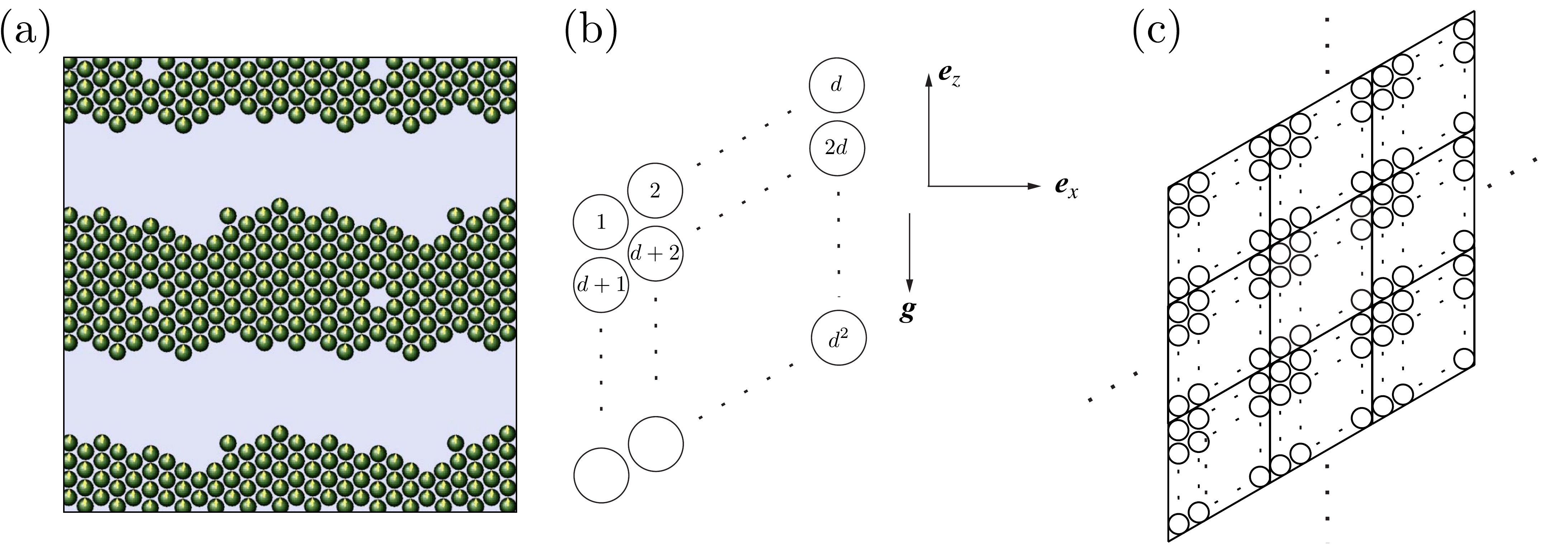}
    \caption{(a) Boundary element simulations \cite{Ishikawa2006,Ishikawa2008a} for $G_{bh}=100$ and $\beta=1$ show the development of stable monolayers of bottom-heavy squirmers, with an equilibrium spacing $\epsilon_0 = 0.002$. (b) Diagram showing the domain of spherical squirmers. The numbering scheme for a $d \times d$ diamond cell is shown. (c) The domain is subject to periodic boundary conditions, as shown. The direction in which the force of gravity acts is denoted by the vector $\bm{g}$.}
    \label{diamond_domain}
  \end{center}
\end{figure}

The present formulation facilitates calculation of the linear and angular velocities of all squirmers in any configuration where lubrication forces dominate. The consequences of perturbing a uniform monolayer of squirmers, subject to periodic boundary conditions, will now be explored. Only cells whose corresponding squirming sets are independent of time ($B_n(t) = B_n \; \forall \; n$. See Eq.~\eqref{squirmer_BC}) will be studied. Consider the diamond-shaped configuration shown in Fig.~\ref{diamond_domain}(b), in which the equilibrium spacing between any two adjacent squirmers is given by $\epsilon_0 a$. In the equilibrium state, all squirmers have an orientation vector $\bm{e} = \bm{e}_z$. For the time being, the motion of the squirmers is limited to the plane of the monolayer. Moreover, all translational and orientational perturbations are restricted to this plane, giving rise to what is essentially a three-dimensional system (2 translational $+$ 1 rotational).

\subsection{Analytical approach}  \label{analytical_section}

The consequences of perturbing the position and orientation of one squirmer will now be investigated. At this stage, time-evolution of the system will not be studied. The purpose of this section is to analytically address the behavior of the system in the small time limit. This corresponds to constructing and solving the matrix-vector equation in Eq.~\eqref{matrix-vector_equation3} once only. Without loss of generality, the perturbed cell is chosen to be squirmer 1, as depicted in Fig.~\ref{diamond_domain}(b). Let the origin of the coordinate system coincide with the center of this squirmer in its equilibrium position. A translational perturbation is initiated, of magnitude $a \delta$ in the direction $\phi$, such that the position vector of the squirmer is given by
\begin{align}
\bm{r} = a \delta \ ( \sin \phi \ \bm{e}_x + \cos \phi \ \bm{e}_z ), \quad \delta \ll 1.
\end{align}
The orientation of the squirmer is also perturbed by $\zeta$, so that
\begin{align}
\bm{e} = \sin \zeta \bm{e}_x + \cos \zeta \bm{e}_z, \quad \zeta \ll 1.
\end{align}
The matrix $\textbf{M}$ and vector $\bm{R}$ depend on the small parameters $\delta$ and $\zeta$. A solution of the following form is sought
\begin{equation}
\bm{x} = \bm{x}_0 + \zeta \bm{x}_1^{\text{r}} + \delta \bm{x}_1^{\text{t}} + \ldots \label{solution_form}
\end{equation}
where the superscripts `r' and `t' represent rotation and translation respectively. Other components of the matrix system can be linearized in the same way:
\begin{align}
\textbf{M} &= \textbf{M}_0 + \zeta \textbf{M}_1^{\text{r}} + \delta \textbf{M}_1^{\text{t}} + \ldots \\
\bm{R} &= \bm{R}_0 + \zeta \bm{R}_1^{\text{r}} + \delta \bm{R}_1^{\text{t}} + \ldots
\end{align}
These expressions are substituted into the original matrix-vector equation Eq.~\eqref{matrix-vector_equation3}, and various orders of $\zeta$ and $\delta$ are equated. The vector $\bm{R}_0$ corresponds to the equilibrium configuration and is equal to $\bm{0}$. It follows that the leading-order solution is $\bm{x}_0 = \bm{0}$. As one might expect, a suspension of evenly spaced squirmers, all pointing in the $z$-direction, do not experience a net force or torque. With this in mind, it is found that
\begin{equation}
\textbf{M}_0 \cdot \bm{x}_1^r = \bm{R}_1^r \qquad \text{and} \qquad \textbf{M}_0 \cdot \bm{x}_1^t = \bm{R}_1^t \label{equate_order_1_short}.
\end{equation}
Recall that the matrix $\textbf{M}$ depends only on the positions of the individuals cells. Thus, in the equilibrium configuration, $\textbf{M}_0$ depends only on the scaled equilibrium spacing, $\epsilon_0$. For a given suspension, the value of $\epsilon_0$ will be known. As such, $\textbf{M}_0$ can be constructed and inverted without knowing anything about the squirming parameters. The vectors $\bm{R}_1^r$ and $\bm{R}_1^t$ can be subsequently constructed. By definition, $\bm{R}_1^r$ must be independent of $\delta$ and thus $\phi$. From the form of $\bm{R}_1^r$, it follows that the elements of the solution $\bm{x}_1^r$ must be of the form
\begin{equation}
x_1^{r,i} = a_i G_{bh} + b_i.
\end{equation}
The matrix $\textbf{M}_0$ cannot be inverted without assuming a particular value of $\epsilon_0$. As such, the coefficients in the above equation must be numerically fitted. The form of the solution arising through small perturbations in the position of the squirmer will now be examined. By definition, $\bm{R}_1^t$ is independent of $\zeta$ and thus $G_{bh}$. Terms in $\bm{R}_1^t$ involve either $\sin(\phi-\phi_0)$ or $\cos(\phi-\phi_0)$ for some $\phi_0$. Since $\textbf{M}_0$ is independent of $\phi$, a solution of the following form is sought:
\begin{equation}
x_1^{t,i} = c_i \sin(\phi - d_i) + e_i.
\end{equation}

\subsubsection{No repulsive force} \label{analytics_norep}

\begin{figure}[h!]
\begin{center}
\includegraphics[width=\textwidth]{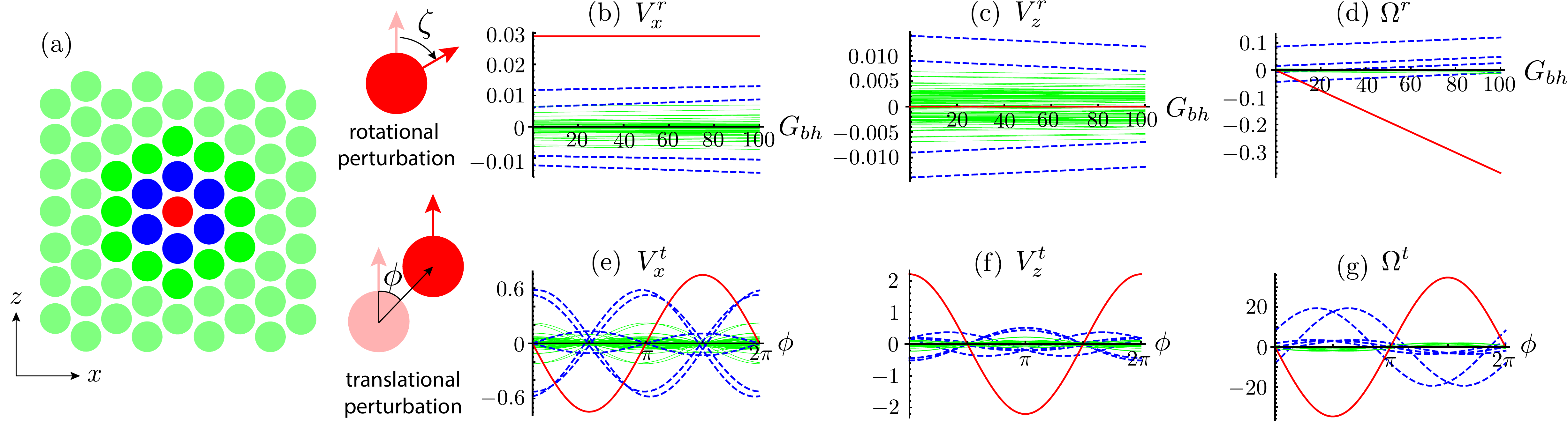}
\caption{Plots showing the linearized solutions associated with all squirmers in the $10 \times 10$ configuration. The results for the perturbed squirmer and its 6 nearest neighbors are shown in red and blue respectively. The green curves correspond to all other squirmers. Results have been computed with $\epsilon_0 = 0.002$ and $\beta = 1$.}
\label{linearised_solutions_norep_Sq_1}
\end{center}
\end{figure}

Consider the situation in which the repulsive force between adjacent squirmers is absent. This is achieved by setting the value of $\kappa_1$ in Eq.~\eqref{repulsive_force} to be equal to zero. For a given $\epsilon_0$ and set of squirming parameters, the leading-order solution in the form of Eq.~\eqref{solution_form} is readily found:
\begin{equation}
V_x =  V_x^r \zeta + V_x^t \delta,\qquad 
V_z =  V_z^r \zeta + V_z^t \delta, \qquad
\Omega = \Omega^r \zeta + \Omega^t \delta. \label{reconstructed_norep}
\end{equation}
The term $\Omega^r$ represents the restoring effect that gravity has on the orientation of the squirmer, and is found to be directly proportional to $G_{bh}$ for the perturbed squirmer. When $\delta = 0$ (ie. no translational perturbation), the orientation will be restored if $G_{bh} > 0$ (since $\Omega^r <0$). For larger $\delta$, a correspondingly larger value of $G_{bh}$ is required to ensure that small perturbations to the orientation decay. In fact, the critical value of $G_{bh}$ is given by
\begin{equation}
G_{bh}^{\text{critical}} = k \times \frac{\delta}{\zeta}, \label{Gbh_crit}
\end{equation}
for some $k>0$. The ratio of the first two squirming modes, defined in Eq.~\eqref{squirmer_BC}, is given by
\begin{equation}
\beta = \frac{B_2}{B_1}. \label{define_squirmratio}
\end{equation}
The parameter $B_2$ is proportional to the stresslet of the squirmer, so $\beta < 0$ and $\beta > 0$ represent pushers and pullers respectively. Perturbing one squirmer in the configuration, will in general, affect all squirmers in the monolayer. Figure~\ref{linearised_solutions_norep_Sq_1} summarizes the results for $\beta=1$. The central squirmer (red) is given either a rotational or a translational perturbation, as shown in Figure~\ref{linearised_solutions_norep_Sq_1}(a). The subsequent linear and angular velocities of all squirmers are then shown (Figs.~\ref{linearised_solutions_norep_Sq_1}(b-d) and Figs.~\ref{linearised_solutions_norep_Sq_1}(e-g) respectively). In each case, the blue curves represent the solutions for squirmers adjacent to the perturbed cell, and green represents the remaining cells in the monolayer. In the case of a rotational perturbation, the central squirmer will experience $V_x^r>0$ (Fig.~\ref{linearised_solutions_norep_Sq_1}(b)), indicating translational instability. The orientational perturbation will decay for that squirmer, but destabilizes the surrounding cells (Fig.~\ref{linearised_solutions_norep_Sq_1}(d)). Translational perturbations in the $x$ and $z$ directions will decay and grow respectively (see Figs.~\ref{linearised_solutions_norep_Sq_1}(e-f)), but at the same time, will destabilize the orientation of the central squirmer (Fig.~\ref{linearised_solutions_norep_Sq_1}(g))). Taken together, these results demonstrate that any perturbations to the uniformly spaced planar array will be linearly unstable, with rotational perturbations causing translational instability, and vice versa. Rotating the cell clockwise or anticlockwise will cause it to move right or left, respectively. Similarly, translating the squirmer right or left will cause it to move clockwise or anticlockwise, respectively.

\begin{figure}[h!]
\begin{center}
\includegraphics[width=0.95\textwidth]{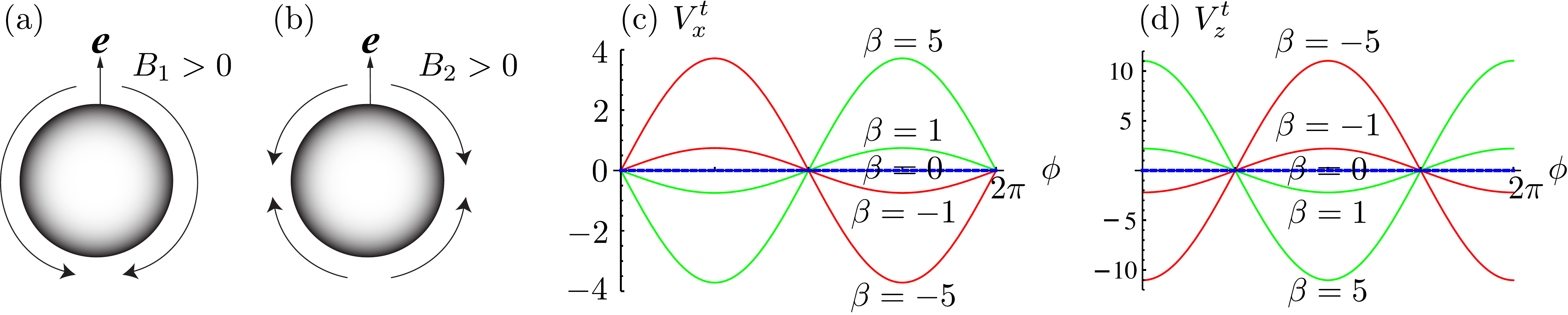}	
\caption{Direction of the tangential velocity at the surface of the squirmer, associated with the first two modes. Results are shown for (a) $B_1>0$ and (b) $B_2>0$. The boundary condition is a superposition of these modes, given by $u_{\theta} \big|_{r=a} = B_1 \sin \theta + B_2 \sin \theta \cos \theta$. (c-d) Plots showing the linearized solutions associated with the perturbed squirmer, for several different values of $\beta$. Results have been computed for a $10 \times 10$ diamond configuration with an equilibrium spacing of $\epsilon_0 = 0.002$. Results are shown for $\beta=0$, $\pm 1$ and $\pm 5$. Positive and negative values of $\beta$ are shown in green and red respectively and the blue curves correspond to $\beta=0$.}  	
\label{modes_V_trans}
\end{center}
\end{figure}

For the case $\beta = -1$, the results (not shown) are extremely similar to those presented in Fig.~\ref{linearised_solutions_norep_Sq_1}. However, the sign of the red curves in panels (e) and (f) is reversed. In order to understand these results, it is helpful reconsider the mechanisms through which the squirming occurs. Figure~\ref{modes_V_trans}(a-b) shows the direction of the tangential velocity for the first two modes of squirming. For $B_2>0$, the second squirming mode serves to draw fluid from the poles of the squirmer ($\theta = 0, \pi$) to the equator ($\theta = \pi/2$). When the position of the squirmer is perturbed in the $x$-direction, this mode restores the position of the squirmer. Conversely, for perturbations in the $z$-direction, this active drawing of fluid away from the poles results in further destabilization from the equilibrium position. The results are reversed for $B_2<0$ (and therefore $\beta<0$).

The functions $V_x^r$, $V_z^r$, $\Omega^r$ and $\Omega^t$ associated with the perturbed squirmer do not vary with $\beta$. Any changes in the value of $\beta$ are manifested only in $V_x^t$ and $V_z^t$, the linear velocities associated with translational perturbations. Consider the plots in Fig.~\ref{modes_V_trans}(c-d), which show these quantities for several different values of $\beta$. The linear velocity of the squirmer after a translational perturbation is directly proportional to $\beta$ (see red curves). It is emphasized again, that the angular velocity of the perturbed squirmer is independent of $\beta$. These observations however, are not in general true for the rest of the squirmers in the configuration.

Recall that there exists a critical value of $G_{bh}$, above which perturbations to the orientation of the squirmer will decay, regardless of the direction, $\phi$, of the translational perturbation. This critical value was shown to depend only on $\Omega^r$ and $\Omega^t$ (see Eq.~\eqref{Gbh_crit}). It has just been found that these two functions associated with the perturbed squirmer are independent of the value of $\beta$ used. It thus follows that $G_{bh}^{\text{critical}}$ does not depend on the ratio $\beta = B_2/B_1$ of the squirming velocities. Importantly, the results obtained in this section are applicable only in the small time limit. The linear and angular velocities have been analyzed for particular squirmer configurations, but the time dependence of the problem has not yet been considered.

\subsubsection{Repulsive force present} \label{analytics_rep}

\begin{figure}[h!]
  \begin{center}
    \includegraphics[width=0.95\textwidth]{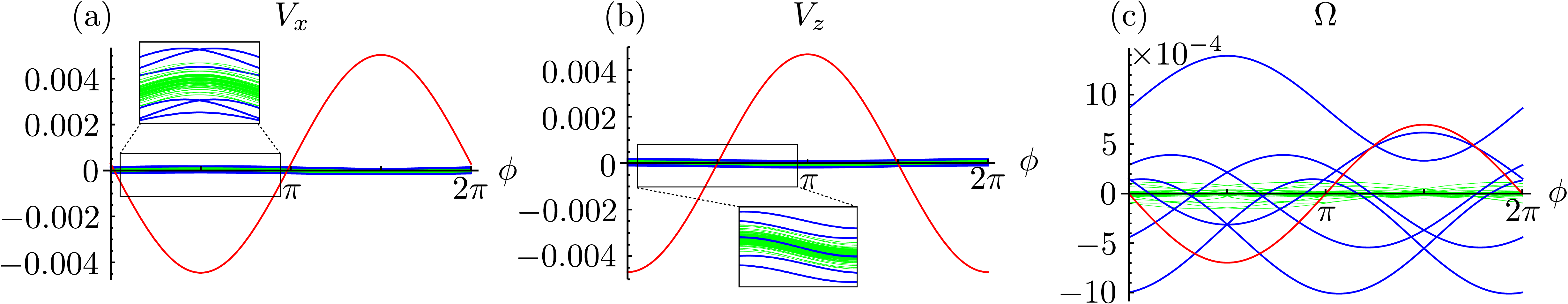}
    \caption{Linearized solutions with repulsive force present and $\beta=1$. Plots showing the linearized solutions associated with all squirmers in the $10 \times 10$ configuration. The results for the perturbed squirmer and its 6 nearest neighbors are shown in red and blue respectively. The green curves correspond to all other squirmers. Results have been computed with $\epsilon_0 = 0.002$, $\zeta = 0.01$, $\delta = \epsilon_0/100$, $\kappa_1=1$, $\kappa_2=10^3$ and $\beta = 1$. $G_{bh}=0$.}
  \label{linearised_solutions_rep_Gbh_0_and_50}
  \end{center}
\end{figure}

In the previous section it was found that perturbations in the position of the squirmer could be either unstable or stable, depending upon the direction of the perturbation $\phi$, and the squirming parameters. The previous analysis will now be repeated, but with the repulsive force outlined in Eq.~\eqref{repulsive_force} included. The velocities in the $x$ and $z$-directions can again be found, as well as the angular velocities for all of the squirmers. Figure \ref{linearised_solutions_rep_Gbh_0_and_50} shows the linearized solutions for a $10 \times 10$ configuration of squirmers. The reconstructed solutions $V_x$, $V_z$ and $\Omega$ for particular values of $\zeta$ and $\delta$ have been plotted, to demonstrate the significance of the repulsive force. It is evident that the position of the central squirmer will be stable subject to small perturbations in either the $x$ or $z$-directions. However, the functions $\Omega^r$ and $\Omega^t$ have not changed upon inclusion of the repulsive force. Consequently, the critical value of $G_{bh}$ required to eliminate small perturbations in $\zeta$ remains the same. That is, $G_{bh}^{\text{critical}}$ is independent of both the value of $\beta$ and the presence of the repulsive force.

\subsection{Numerical approach} \label{numerical_section}

In Section~\ref{analytical_section}, the analytical form of the linear and angular velocities associated with squirmers in a large uniform suspension were studied. In particular, the case where one squirmer was subject to small translational and angular perturbations was considered. However, these results were not able to address the long-term behavior of the suspension following a perturbation from the equilibrium. A numerical study into the dynamics of the monolayer will now be undertaken, using the formulation presented in Section~\ref{lubrication_multiple_spheres}. The configuration is the same as the one presented in Fig.~\ref{diamond_domain}(b) and is again subject to periodic boundary conditions, as depicted in Fig.~\ref{diamond_domain}(c). 

\begin{figure}[h!]
  \begin{center}
  \includegraphics[width=\textwidth]{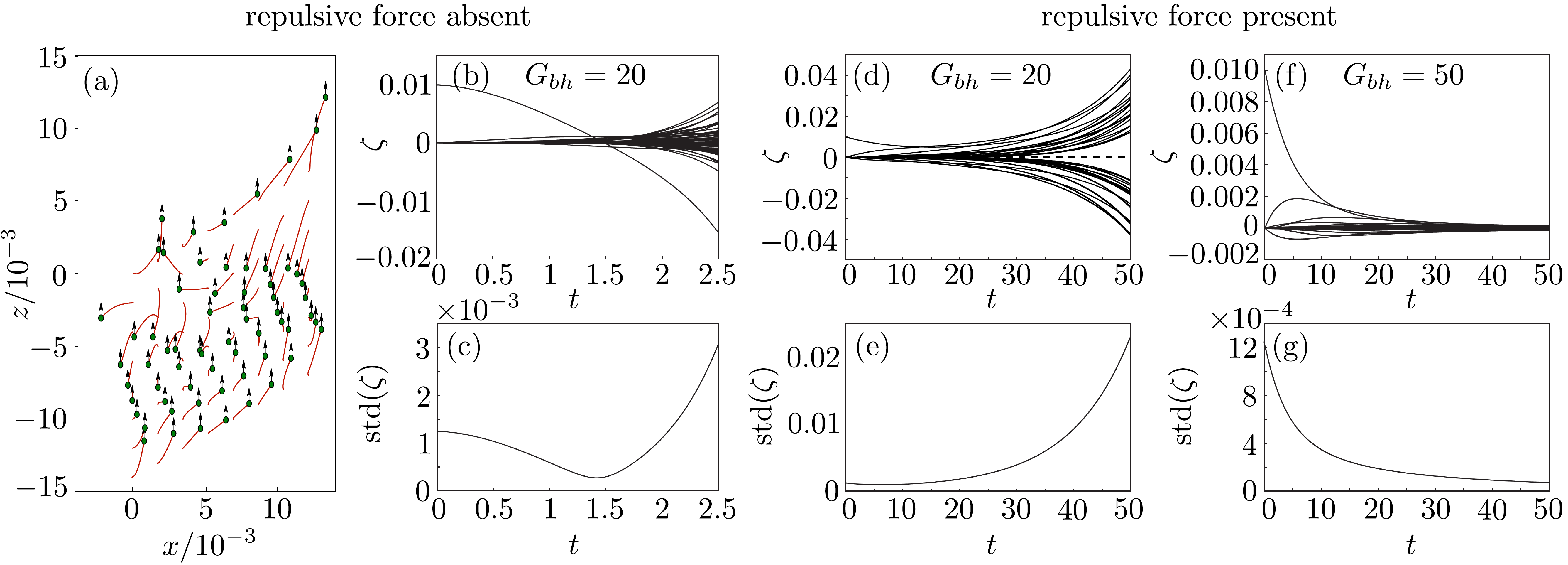}
     \caption{One single squirmer is perturbed in an otherwise uniform monolayer. (a) Figure showing the trajectories of all 64 squirmers over the course of the simulation. For the purposes of plotting, the positions have been scaled so that the radius of each squirmer is 0, even though the simulation was conducted with $a=1$. (b) Plots showing the orientation $\zeta$ and (c) corresponding standard deviation of all squirmers as a function of time. Parameters used are $\epsilon_0=2 \times 10^{-3}$, $\beta=1$, $\kappa_1=0$, $\phi = 3\pi/2$, $\delta = \epsilon_0/1000$, $\zeta=1/100$ and $G_{bh}=20$. 
    Simulations were repeated with repulsive force present ($\kappa_1=1$) for (d-e) $G_{bh}=20$ and (f-g) $G_{bh}=50$. Under these conditions, the lattice is translationally stable, but the squirmers still require sufficiently large value of $G_{bh}$ for orientational stability.}
    \label{numerical_single_pert}
  \end{center}
\end{figure}

To begin with, the consequences of perturbing one squirmer in the uniform monolayer will be explored. For small perturbations, this system was studied analytically in Section~\ref{analytical_section}. It was found that the position of the squirmer was stable, provided the repulsive force between adjacent squirmers was included. It was also found that there exists a critical value of $G_{bh}$, above which perturbations to the orientation of the squirmer will decay, regardless of the direction, $\phi$, of the translational perturbation. Recall that this value was independent of the presence of the repulsive force. The functional form of this critical value is given in Eq.~\eqref{Gbh_crit}, and was derived under the assumption that the neighboring squirmers were all left unperturbed. The results in Fig.~\ref{numerical_single_pert} show the consequences of perturbing one squirmer in an $8 \times 8$ system. The equilibrium spacing is again considered to be $\epsilon_0=2 \times 10^{-3}$ and the perturbation is given by $\zeta = 1/100$ and $\delta = \epsilon_0/1000$ with $\phi = 3 \pi/2$. From the analysis in Section~\ref{analytical_section}, it is known that for these parameters, to ensure $\Omega < 0$ for the perturbed squirmer requires $G_{bh} > 1.84$. The value $G_{bh}=20$ is used, which is known to be well beyond this critical value.

From the previous analytical work, it is known that the value of $G_{bh}$ used here guarantees $\Omega <0$ for the first time-step. However, the only way the long-time behavior can be assessed is through these numerical simulations. For $t \ll 1$, the system without intercellular repulsive force ($\kappa_1=0$) appears to be stable, with the orientation of the perturbed squirmer beginning to be restored (Fig.~\ref{numerical_single_pert}(b)). However, as time progresses, the other squirmers in the system begin to move (Fig.~\ref{numerical_single_pert}(a)). Indeed, the system becomes unstable as time progresses, with the orientation of all squirmers growing in magnitude. The perturbations that develop in the surrounding squirmers will act to destabilize the central squirmer.

It was found earlier that in the absence of the repulsive force, for $\beta > 0$, each squirmer is stable and unstable to translational perturbations in the $x$ and $z$-directions respectively, with the converse true for $\beta < 0$. Figure~\ref{numerical_single_pert} demonstrates this phenomenon clearly, with the squirmers drifting towards one another in the $z$-direction. The repulsive force outlined in Section~\ref{analytical_section} is now reinstated. This prevents the squirmers from coming too close together, since translational perturbations are quickly eliminated. Consider the plots in Fig.~\ref{numerical_single_pert}(d-g), which show the orientation of 64 steady squirmers over the interval $t \in [0,50]$, for two different values of $G_{bh}$. Even with the orientation of only one single squirmer perturbed, the whole system eventually becomes unstable for $G_{bh}=20$. However, the system is stable for large $t$ when $G_{bh}=50$. 

Since the repulsive force quickly restores the position of the squirmers to their equilibrium value, where $\delta \rightarrow 0$, the value of $G_{bh}$ used is expected to be well above the critical value derived earlier. Nevertheless, instability is observed among the orientation of the squirmers. In the early stages of the simulation, the orientations of the neighbors become perturbed, causing the original perturbed squirmer to become further destabilized. This results in an increase in the corresponding value of $G_{bh}$ required to eliminate all angular perturbations. Since the restoring force quickly returns the squirmers to their equilibrium positions, the critical value of $G_{bh}$ required for angular stability does not depend strongly on the translational perturbations initially given to the squirmers. The stability of the orientation of the squirmers is dictated by the angular velocities associated with small perturbations to the orientations rather than positions. The interactions between $\Omega^r$ for various squirmers are key in determining the critical value of $G_{bh}$ required for stability. In addition to the fact that perturbations to the orientations grow when $G_{bh} < G_{bh}^{\text{critical}}$, another interesting feature of Fig.~\ref{numerical_single_pert}(d) is the splitting of these orientations in a dichotomous fashion. As time progresses, the squirmers rotate away from vertical in a coordinated manner. It will be shown later that this phenomenon also occurs in other configurations in which instabilities develop.
 
\begin{figure}[htp]
\begin{center}
\includegraphics[width=0.9\textwidth]{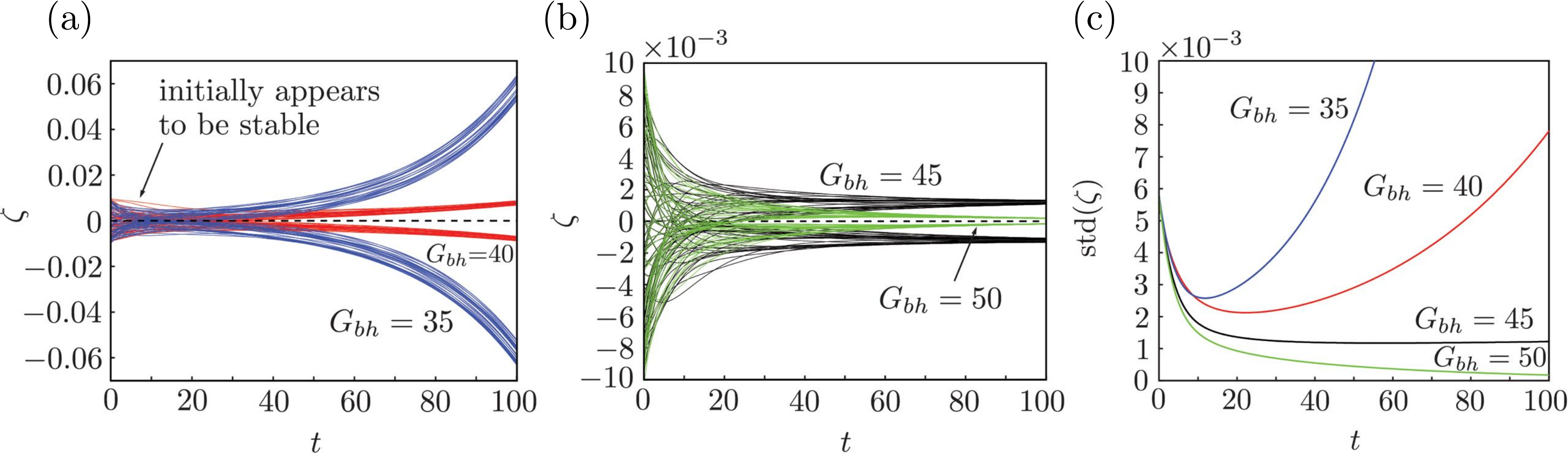}
\caption[Orientation of squirmers as a function of time for various values of $G_{bh}$]{Orientation of every squirmer in an $8 \times 8$ diamond, as a function of time. In each simulation, every squirmer is initially given a random perturbation to its position and orientation, with amplitude $\delta = \epsilon_0/100$ and $\zeta = 1/100$ respectively. Results have been computed for $G_{bh}=35$, 40, 45 and 50 over the interval $t \in [0,100]$. Also shown is the standard deviation of $\zeta$ as a function of time. Additional parameters used are given by $\epsilon_0=2 \times 10^{-3}$, $\kappa_1=1$, $\kappa_2=10^3$ and $\beta=1$.}
\label{zeta_vs_time_rep}
\end{center}
\end{figure}

 We investigated the effects of perturbing the position and orientation of all squirmers in the uniform monolayer. The squirmers were given a random perturbation to both their orientation and position, with amplitudes $\zeta = 1/100$ and $\delta = \epsilon_0/100$ respectively. The stabilizing effect that gravity has on the suspension is evident (see Fig.~\ref{zeta_vs_time_rep}). For $G_{bh}=35$ and 40 the system is unstable, while for $G_{bh}=50$ the system is stable for large $t$. Although not shown here, the results for $G_{bh}=0$ yield $\text{std}(\zeta) \rightarrow \pi / \sqrt{3}$ for large $t$, corresponding to a uniform distribution in which there is no preferred orientation.

\section{Monolayer of squirmers between vertical rigid walls} \label{section_twowalls}

In the preceding sections, the positions of the squirmers were restricted to lie in the $x$-$z$ plane. This condition can be relaxed in order to permit out of plane motion. Under these conditions, the monolayer is unstable, with perturbations in the $y$-direction growing (see Supplementary Information Section~S2 for detailed analysis).
However, a uniform monolayer of squirmers in an otherwise unbounded fluid is, in any case, an unrealistic situation. One method of maintaining a monolayer of spherical squirmers is to use a Hele-Shaw cell which is sufficiently thin. To this point, the forces and torques on the squirmers arising due to either sphere-sphere interactions or the effects of gravity have been considered. It is straightforward to extend to the case where the uniform monolayer of steady, spherical squirmers is situated between two plane parallel walls. The two planes are defined by $y = \pm a(1+\epsilon_0^{\text{wall}})$ so that the minimum clearance between the squirmers and the wall in the equilibrium configuration is $\epsilon_0^{\text{wall}} a$. Gravity is still considered to act in the negative $z$-direction. 

As the lubrication analysis presented earlier applies to two spheres, each of arbitrary radius, the forces and torques acting on a squirmer interacting with the planes can easily be found by taking the limit $\lambda \rightarrow \infty$. The forces acting on the squirmers due to their translational and rotational motion must also be considered. For this, the results presented in ref \cite{KimAndKarrila:Microhydrodynamics} are applied. A short-range repulsive force between the spheres and the walls is also incorporated into the model, as in Eq.~\eqref{repulsive_force}, with the parameters $\kappa_1^{\text{wall}}$ and $\kappa_2^{\text{wall}}$. Equation~\eqref{all_forces_and_torques} is modified by the inclusion of extra terms to account for the walls (see SI Eq.~\eqref{all_forces_and_torques_wall}).

To explore the influence that squirming strength and bottom heaviness have on the monolayer stability, we performed 702 simulations across a range of values for $G_{bh}$ and $\beta = B_2/B_1$ (see Fig.~\ref{modes_V_trans}(a-b) for schematic). This enables us to explicitly investigate the differences between pushers ($\beta <0$) and pullers ($\beta > 0$). In each simulation, the squirmers' orientations were subject to random initial conditions, and the long time dynamics were observed. The effect of the repulsive force is to stabilize the positions of the squirmers, retaining the lattice like structure. It is therefore sufficient to consider the orientation from vertical, $\theta_i$, of squirmers in each monolayer. Several qualitatively different dynamics emerge, depending on the parameter combination $(\beta , G_{bh})$. 

\begin{figure}[htp!]
\begin{center}
\includegraphics[width=0.65\textwidth]{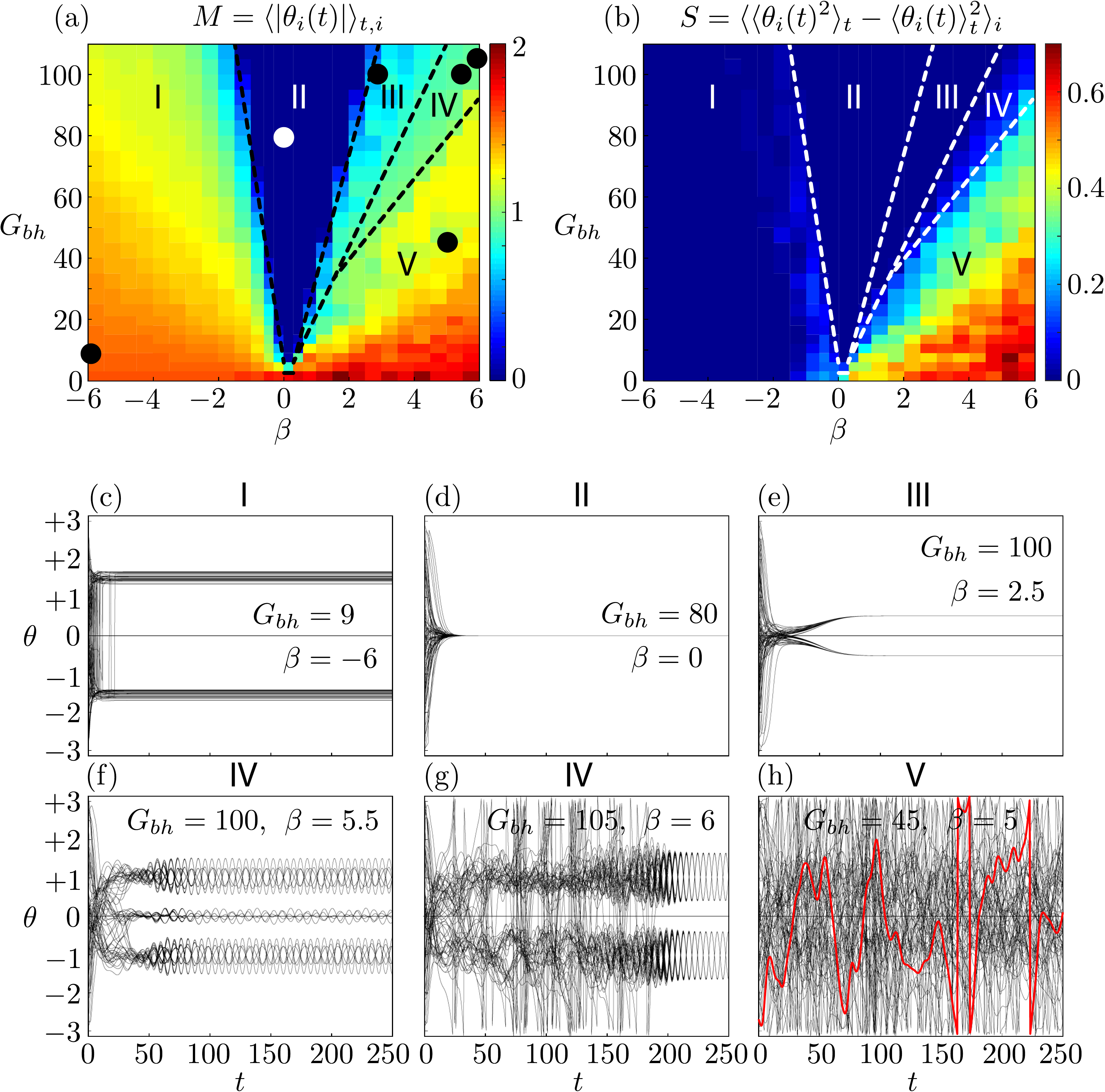}
\caption{Lubrication simulations of a squirmer monolayer situated between two plane parallel walls. (a) The mean angle from vertical, $M$, and (b) average variance for each squirmer, $S$, are shown across a broad range of $G_{bh}$ and $\beta$ values. (c-h) Representative results from different regions of parameter space highlight the qualitatively different long term dynamics. Parameters used include $\epsilon_0=\epsilon_0^{\text{wall}} = 2 \times 10^{-3}$, $\kappa_1 = \kappa_1^{\text{wall}}=1$, $\kappa_2= \kappa_2^{\text{wall}} =10^3$. The parameter combinations in c-h are depicted as circles in a. }
\label{parameter_search}
\end{center}
\end{figure}

In order to quantify the dynamics for various parameter choices $(\beta , G_{bh})$, we analyze the time-dependent angle $\{ \theta_i (t) \}$ for all squirmers $i=1,\ldots,N$ in a given simulation. Firstly, we define $M = \langle |\theta_i(t)| \rangle_{t,i}$, averaged over time and all squirmers in the monolayer. The parameter $M=0$ if and only if all squirmers converge to a vertical orientation at large $t$. However, in the case of $M \neq 0$, this parameter is unable to distinguish between steady states (e.g. Fig.~\ref{parameter_search}(e)) and chaotic results (Fig.~\ref{parameter_search}(h)). We therefore define a second parameter, $S$, calculated by taking the variance of each time-dependent signal $\theta_i(t)$, and subsequently averaging over the squirmer population.
\begin{equation}
S = \langle \langle \theta_i(t)^2 \rangle_t - \langle \theta_i(t) \rangle_t ^2 \rangle_i.
\end{equation}
The parameter $S$ will be zero if every squirmer converges to a constant orientation, regardless of its value. This parameter therefore provides great utility in distinguishing between equilibrium structures and other results. For each of the 702 simulations, the parameters $M$ and $S$ were calculated, the results of which are summarized in Fig.~\ref{parameter_search}(a) and Fig.~\ref{parameter_search}(b) respectively. 

Across the range of parameters studied, 5 different characteristic behaviors were observed for $t \gg 1$. The simplest case is that in which all squirmers eventually orient vertically, $\theta_i \rightarrow 0$ (Case II in Fig.~\ref{parameter_search}, see for example Fig.~\ref{parameter_search}(d)), and is identified as when both $M=0$ and $S=0$. This is precisely the equilibrium structure initially observed by Ishikawa {\it et al.} \cite{Ishikawa2008a, Ishikawa2008} for $\beta=1$, $G_{bh}=100$, and investigated analytically in Section~\ref{analytical_section}. From the random initial conditions studied here, the system can converge to this vertical state for either pushers ($\beta <0$) or pullers ($\beta > 0$), provided $G_{bh}$ is sufficiently large. For pushers ($\beta <0$), only one other type of behavior is possible, in which all squirmers converge to a non-zero equilibrium orientation (Case I in Fig.~\ref{parameter_search}, see for example Fig.~\ref{parameter_search}(c)). These dynamics occur when the second-order squirming mode is large enough to destabilize the vertically oriented monolayer. 

For pullers ($\beta > 0$), a richer set of dynamics is possible. For a given value of $G_{bh}$, increasing $\beta$ beyond a critical value results in an abrupt transition from $M=0$ to $M>0$. The system adopts a bistable state, in which squirmers possess a finite and constant tilt angle (Case III in Fig.~\ref{parameter_search}, see Fig.~\ref{parameter_search}(e)), qualitatively similar to Case I. Increasing $\beta$ further results in oscillations about these values (Case IV, see also Fig.~\ref{parameter_search}(f,g)). For sufficiently large $\beta$, the entire system becomes unstable (Case V). Figure~\ref{parameter_search}(h) illustrates these unstable dynamics, with the orientation of one squirmer shown in red.

\section{Investigation of tilted structures and oscillatory states} \label{tilted_structures}

\begin{figure}[h!]
\begin{center}
\includegraphics[width=0.65\textwidth]{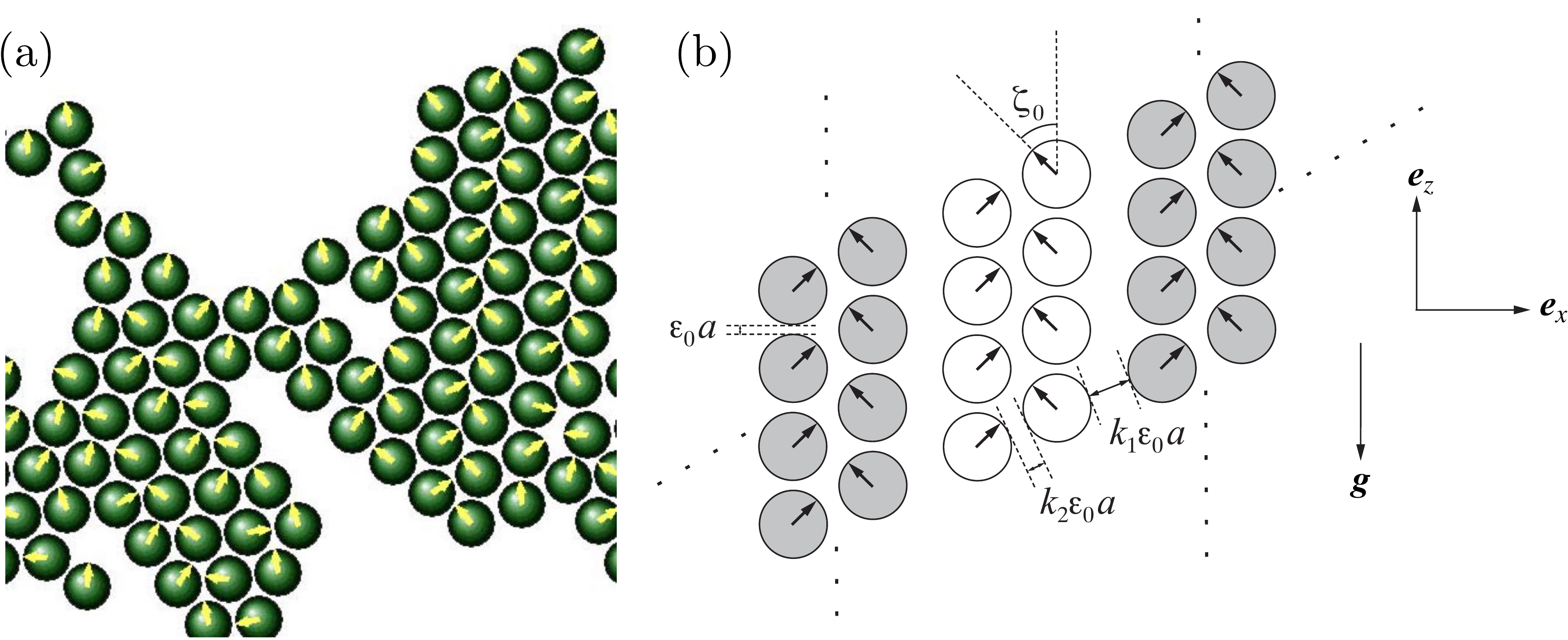}
\caption[Coherent structures containing tilted spherical squirmers]{(a) Depiction of coherent structures (Ishikawa \textit{et al.}) which have formed within a monolayer of spherical squirmers. Results have been computed using $\beta=5$ and $G_{bh}=100$. (b) Schematic diagram showing this equilibrium configuration in which all squirmers are oriented at an angle of $\zeta = \pm \zeta_0$ from vertical. In this configuration, three different equilibrium spacings $\epsilon_0 a$, $k_1 \epsilon_0 a$ and $k_2 \epsilon_0 a$ are permitted.}
\label{tilted_stable_structure}
\end{center}
\end{figure}

The numerical simulations of Section~\ref{section_twowalls} revealed the existence of stable states in which all squirmers adopt a non-zero mean orientation from vertical, either constant in value or oscillating in time. By symmetry, the configuration in which all squirmers in the lattice are vertically oriented is a steady state, and the linear and angular velocities of all squirmers in the periodic lattice will be zero. However, the conditions under which the ``tilted equilibrium'' can occur are not immediately clear. Ishikawa \textit{et al.} also discovered stable coherent structures in which the squirmers do not orient themselves in a vertical direction, even in the presence of strong bottom-heaviness (see Fig.~\ref{tilted_stable_structure}). The only difference between Figs.~\ref{diamond_domain}(a) and \ref{tilted_stable_structure}(a) is that the value of $\beta$ has been increased from 1 to 5. This corresponds to a shift in parameters equivalent to moving from Case II to Case IV in Fig.~\ref{parameter_search}. The structure in which all squirmers possess some orientation of magnitude $\zeta_0$ from vertical, as shown in Fig.~\ref{tilted_stable_structure}(b), will now be studied. It will be the goal of this section to understand the nature of this equilibrium state.

Until now, the equilibrium spacing between adjacent cells has been considered to be uniform throughout. Three different values are now permitted, namely $\epsilon_0 a$, $k_1 \epsilon_0 a$ and $k_2 \epsilon_0 a$. In the case where $k_1=k_2=1$, it is found that the net force on the squirmers is zero provided $\zeta_0=0$ or $B_1=0$. The former case has already been studied in detail, whilst the latter case involves squirmers that would not be able to swim in an unbounded fluid (swimming speed = $2B_1/3$). If the columns in Fig.~\ref{tilted_stable_structure}(b) are evenly spaced, with $k_1 = k_2 \neq 1$, then the same conditions are required. The equilibrium configuration depicted in Fig.~\ref{tilted_stable_structure}(a) cannot be achieved with $k_1=k_2$ unless $B_1=0$. Such a configuration would, by symmetry, be independent of the inter-particle repulsive force.

\begin{figure}[htp]
\begin{center}
\includegraphics[width=0.65\textwidth]{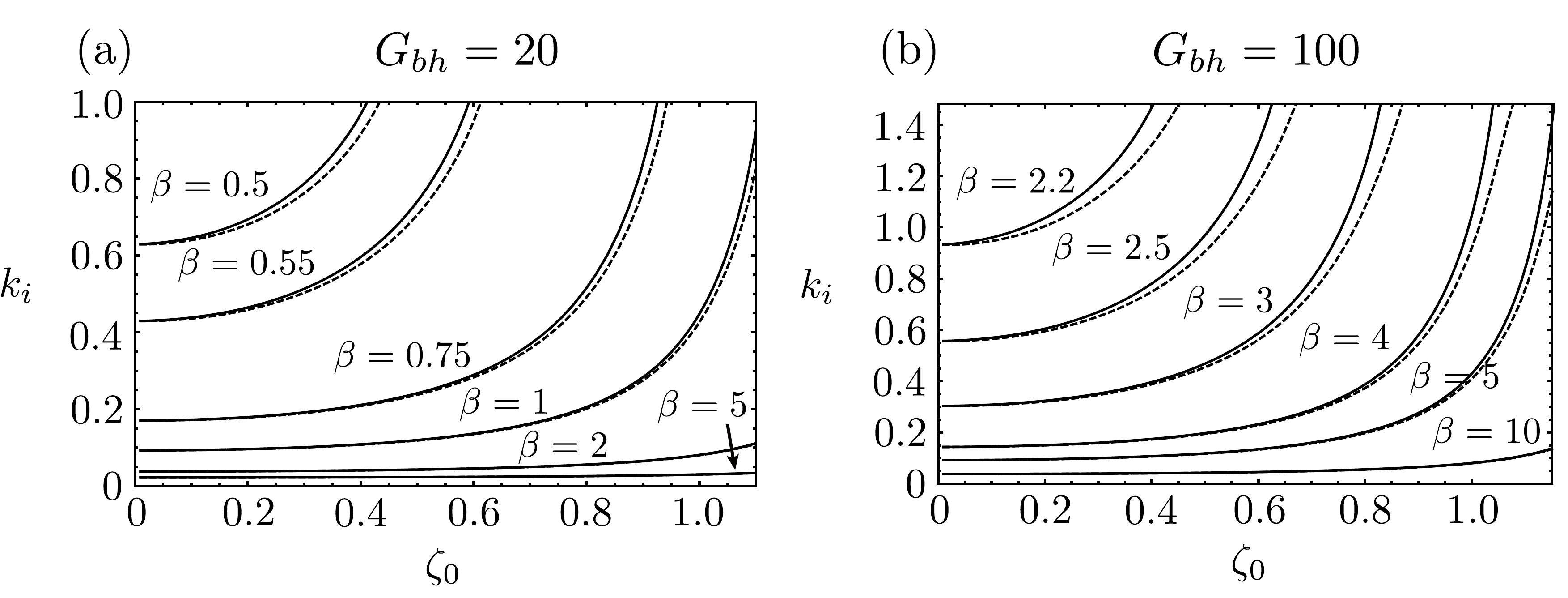}
\caption{Equilibrium spacing in ensemble of uniformly tilted squirmers. Plots showing the values of $k_1$ (smooth) and $k_2$ (dashed) as functions of $\zeta_0$ for (a) $G_{bh}=20$ and (b) $G_{bh}=100$. Results have been computed with $\epsilon_0 = 2/1000$, $\kappa_1=1$ and $\kappa_2 = 10^3$ for various $\beta$.}
\label{tilted_k}
\end{center}
\end{figure}

Consider now the case where $k_1 \neq k_2$. In this scenario, it is immediately obvious that the net force experienced by each squirmer as a consequence of the repulsive force presented in Eq.~\eqref{repulsive_force} will be non-zero, and so, the existence of an equilibrium configuration will depend on the presence of this force. Nevertheless, the analysis is continued in an attempt to account for the results in Fig.~\ref{parameter_search} and boundary element simulations of Ishikawa {\it et al.} (Fig.~\ref{tilted_stable_structure}(a)). By specifying the values of $\epsilon_0$, $\kappa_1$ and $\kappa_2$, it is possible to calculate the values of $k_1$ and $k_2$ for any $\beta$, $G_{bh}$ and equilibrium orientation $\zeta_0$. For a given experiment, $\beta$ and $G_{bh}$ will be known \textit{a priori} and so the spacings $k_1 \epsilon_0 a$ and $k_2 \epsilon_0 a$ will be functions of the equilibrium orientation $\zeta_0$. Figures~\ref{tilted_k}(a) and \ref{tilted_k}(b) show the values of $k_i$ associated with $G_{bh}=20$ and $G_{bh}=100$ respectively. The curve in Fig.~\ref{tilted_k}(b) corresponding to $\beta=5$ incorporates exactly the same parameters as in Fig.~\ref{tilted_stable_structure}(a). In order to determine the equilibrium orientation $\zeta_0$, an additional piece of information is required. It would be possible, for instance, to demand that the mean equilibrium spacing between adjacent cells is equal to $\epsilon_0$. That is, $(k_1+k_2)/2=1$. The corresponding values are given by $k_1=1.07134$, $k_2=0.92866$ and $\zeta_0=1.12638$. This value of $\zeta_0$ is very similar to that observed in Fig.~\ref{tilted_stable_structure}(a). The advantage of choosing values of $k_i$ as close to 1 as possible is that it minimizes the effects associated with the repulsive force. For given values of $\beta$ and $G_{bh}$ it is possible to find the equilibrium spacings $k_1$ and $k_2$, and orientation $\zeta_0$. Although this reveals the existence of a tilted equilibrium configuration, it does not assess the stability of the monolayer in that case. Depending on the parameter configuration $(\beta,G_{bh})$, the monolayer may be bistable (Case III), oscillatory (Case IV), or completely unstable (Case V).

\section{Discussion and Conclusions} \label{discussion}

While fully resolved boundary element simulations of interacting spherical squirmers revealed stable hexagonal lattice configurations \cite{Ishikawa2008a}, the model complexity prevented a simple understanding of the mechanisms behind this stability. Moreover, the computational intensity precluded a broad exploration of parameter space relevant to a range of biological and synthetic micro-swimmers. Here we have developed a semi-analytical framework to predict the dynamics of dense suspensions of spherical squirmers. We began by solving the Stokes equations to second order between closely-separated squirmers. This followed similar steps to ref \cite{Ishikawa2006}, but to higher order, as required to calculate the normal force. These analytical expressions were then utilized in a `lubrication simulation', to assess the global dynamics of a dense monolayer of squirmers. This revealed that pairwise lubrication interactions, in conjunction with a short-range repulsive force, were sufficient to account for the stable states observed in previous studies. This framework therefore provides a computationally inexpensive means of investigating the dynamics of dense suspensions of swimming microorganisms.

Initial studies of the monolayer restricted the motion of the spheres to lie in a plane, even though the fluid was unbounded and three-dimensional. Further analysis of this monolayer confirms the intuitive result that it is unstable subject to small out-of-plane perturbations (see SI Fig.~S1 for further information).
The inclusion of nearby plane parallel walls, as in the case of a rigid Hele-Shaw cell, maintains the structure of the monolayer (see SI Fig.~S2),
with orientational perturbations again eliminated for sufficiently large values of $G_{bh}$. For every value of $\beta$ and $G_{bh}$ studied, suspensions of pushers ($\beta<0$) were stable for large $t$, with all squirmers converging either to vertical, or a finite tilt angle. Conversely, pullers ($\beta>0$) exhibited a range of qualitatively different states (see Fig.~\ref{parameter_search}), with orientations being completely unstable for sufficiently large $\beta$.

In the present framework, we have neglected any density difference between the squirmers and the fluid, which would lead to sedimentation \cite{Drescher:2009vn}. The inclusion of a Stokeslet term would modify the flow through the monolayer, and therefore potentially influence the stability calculations. This is the subject of future work.

The equilibrium spacing between adjacent squirmers, $\epsilon=0.002$, was chosen to match the stable value emerging from full boundary element simulations \cite{Ishikawa2008a}. Under these conditions, the logarithmic singularities (see Eq.~\eqref{forces_and_torques}) dominate the expressions for the hydrodynamic forces and torques. This paper focusses on the collective dynamics of monolayers of spherical squirmers, but the framework could be readily extended to model fully 3D concentrated suspensions. Although the present analysis could in principle also be applied to polydisperse suspensions, it is likely that substantial variations in the separation $\epsilon$ would limit applicability of the lubrication approximations. The colonial alga {\it Volvox carteri} is a very good realization of Lighthill's spherical squirmer \cite{Lighthill:1952} (with $\beta<0$), but there are significant experimental challenges in preparing a monodisperse suspension of {\it Volvox}. Experimental investigation of the present system is therefore most likely to be achieved for large suspensions of identical synthetic microswimmers \cite{Thutupalli2018} situated in a vertical Hele-Shaw cell.

\section*{Acknowledgements}

The authors thank T. Ishikawa, R.E. Goldstein, and M. Polin for useful discussions, and The University of Melbourne's High Performance Computer {\it Spartan}. This work was supported by a Gates Cambridge Scholarship, a Human Frontier Science Program Cross-Disciplinary Fellowship, and a Discovery Early Career Researcher Award DE180100911 (D.R.B.).

%


\setcounter{figure}{0}
\renewcommand{\thefigure}{S\arabic{figure}}
\setcounter{equation}{0}
\renewcommand{\theequation}{S\arabic{equation}}
\setcounter{section}{0}
\renewcommand{\thesection}{S\arabic{section}}

\begin{center}
\begin{LARGE}
{\sc Supplementary Information}
\end{LARGE}
\end{center}

\section{Hydrodynamic interactions between spherical squirmers} \label{SI_interaction_derivation}

This analysis follows similar steps to that presented in ref \cite{Ishikawa2006}, but with some corrections, and completed to higher order. The spatial coordinates are scaled according to $\epsilon^{1/2} a X=x$, $\epsilon^{1/2} a Y=y$, $\epsilon a Z=z$ and thus $\epsilon^{1/2} a \rho=\rho^*$. The boundaries of spheres 1 and 2 within the lubrication region can then be written as follows:
\begin{equation}
Z = H_1 = 1 + \frac{\rho^2}{2} + \mathcal{O}(\epsilon), \qquad Z = H_2 = - \frac{\rho^2}{2 \lambda} + \mathcal{O}(\epsilon), \label{H scaled}
\end{equation}
where $\rho = \sqrt{X^2+Y^2}$. By linearity of the Stokes equations, the problem involving two squirming spheres in a fluid that is at rest infinitely far away can be broken down into two distinct problems. The first has the squirming-sphere boundary condition on sphere 1 and zero velocity boundary condition on sphere 2. The second problem has zero velocity on sphere 1 and the squirming-sphere boundary condition on sphere 2. Only the former problem will be studied, since solving this will immediately yield the solution to the latter. For a solitary squirmer immersed in a fluid which is at rest at infinity, one can express the fluid velocity field as
\begin{align}
\bm{u}_{\text{sol}} &= - \frac{1}{3} \frac{a^3}{r^3} B_1 \bm{e} + B_1 \frac{a^3}{r^3} \frac{\bm{e}\cdot \bm{r}}{r} \frac{\bm{r}}{r} + \sum_{n=2}^{\infty} \Bigg( \frac{a^{n+2}}{r^{n+2}} - \frac{a^n}{r^n} \Bigg) B_n P_n \Big(\frac{\bm{e} \cdot \bm{r}}{r} \Big) \frac{\bm{r}}{r} \nonumber \\
& \qquad + \sum_{n=2}^{\infty} \Bigg( \frac{n}{2} \frac{a^{n+2}}{r^{n+2}} - \Big( \frac{n}{2}-1 \Big) \frac{a^n}{r^n} \Bigg) B_n W_n \Big(\frac{\bm{e} \cdot \bm{r}}{r} \Big) \Big( \frac{\bm{e} \cdot \bm{r}}{r} \frac{\bm{r}}{r} - \bm{e} \Big),
\end{align}
where $\bm{e}$ is the swimming direction, $\bm{r}$ is the position vector and $r = |\bm{r}|$. By writing the position vector of sphere 1 as $\bm{r}_1 = \bm{r} - (1+\epsilon) a \bm{e}_{Z}$ where $\bm{e}_{Z} = (0,0,1)$ and performing a Taylor series expansion of $W_n \big( \tfrac{\bm{e} \cdot \bm{r}_1}{r_1} \big)$ about the point $- \bm{e}_1 \cdot \bm{e}_z$, the fluid boundary condition on the surface of sphere 1 can be written in the form
\begin{equation}
\bm{u} = \bm{u}_A + \epsilon^{1/2} \bm{v}_A + \epsilon \bm{w}_A + \mathcal{O}(\epsilon^{3/2}), \label{u expansion}
\end{equation}
where the functions $\bm{u}_A$, $\bm{v}_A$ and $\bm{w}_A$ are expressed as infinite series over the squirming modes. From the boundary conditions presented in Eq.~\eqref{u expansion}, it seems logical to attempt to express the velocity and pressure as power series in $\epsilon^{1/2}$. Consider the following expansions for the fluid velocity $\bm{u} = (u,v,w)$ and pressure $p$:
\begin{align}
u &= u_0 + \epsilon^{1/2} u_1 + \mathcal{O}(\epsilon), \label{u_expansion} \\
v &= v_0 + \epsilon^{1/2} v_1 + \mathcal{O}(\epsilon), \label{v_expansion} \\
w &= \epsilon^{1/2} w_0 + \epsilon w_1 +  \mathcal{O}(\epsilon^{3/2}), \label{w_expansion} \\
p &= \epsilon^{-3/2} p_0 +  \epsilon^{-1} p_1 + \mathcal{O}(\epsilon^{-1/2}). \label{p_expansion} 
\end{align}
With these in mind, the $x$, $y$ and $z$-components of the Stokes equations to various orders in $\epsilon$ can be extracted.

\subsubsection{The first-order solution}

From the leading-order $z$-component of the Stokes equations, it is evident that $p_0 = p_0(X,Y)$. Furthermore, on spheres 1 and 2, the fluid velocity is given by $\bm{u}_A$ and $\bm{0}$ respectively. Subject to these boundary conditions, the leading-order $x$ and $y$-components of the Stokes equations are integrated to find the following expressions:
\begin{align}
u_0 &= \frac{1}{2} \frac{a}{\mu} \frac{\partial p_0}{\partial X} (Z-H_1)(Z-H_2) + \frac{Z-H_2}{H} \bm{u}_A \cdot \bm{e}_x, \label{u_0_solution} \\
v_0 &= \frac{1}{2} \frac{a}{\mu} \frac{\partial p_0}{\partial Y} (Z-H_1)(Z-H_2) + \frac{Z-H_2}{H} \bm{u}_A \cdot \bm{e}_y, \label{v_0_solution}
\end{align}
where $H = H_1 - H_2 = 1 + \frac{\lambda+1}{2 \lambda} \rho^2 + \mathcal{O}(\epsilon)$. Integrating the leading-order component of the continuity equation and utilizing Eqs.~\eqref{u_0_solution} and \eqref{v_0_solution} yields the Reynolds equation:
\begin{equation}
\frac{H^3}{12} \nabla_{\perp}^2 p_0 + \frac{H^2}{4} \bigg( \frac{\lambda+1}{\lambda} \bigg) \bm{\rho} \cdot \nabla_{\perp} p_0 + \frac{\mu}{a} \bigg( \frac{\lambda+1}{2 \lambda} \bigg) \sum_n B_n W_n \big( - \bm{e} \cdot \bm{e}_z \big) \bm{e} \cdot \bm{\rho} = 0, \label{Reynolds_eqn_0}
\end{equation}
where we have made use of the fact that $\bm{\rho} \cdot \bm{u}_A = - \sum_n B_n W_n \big( - \bm{e} \cdot \bm{e}_z \big) \bm{e} \cdot \bm{\rho}$. Recall that $p_0=p_0(X,Y)$, or equivalently $p_0(\rho,\phi)$. A solution to Eq.~\eqref{Reynolds_eqn_0} of the form $p_0 (\rho,\phi) = q_0 (\rho) \bm{e} \cdot \bm{e}_{\rho}$ is found, where
\begin{equation}
q_0 (\rho) = Q_0(\rho) \sum_n B_n W_n \big( -\bm{e} \cdot \bm{e}_z \big) \quad \text{and} \quad Q_0(\rho) = \frac{6 \mu}{5 a} \frac{\rho}{H^2}. \label{leading_order_pressure}
\end{equation}
Figure~\ref{pressure_first_and_second_order}(a) shows how $Q_0(\rho)$ varies as a function of $\rho$ for various values of $\lambda$.

\subsubsection{The second-order solution}

In a fashion completely analogous to the first-order case, the second-order fluid velocities can be derived. These are equivalent to the expressions in Eqs.~\eqref{u_0_solution} and \eqref{v_0_solution} but with $\bm{v}_A$ replacing $\bm{u}_A$.
\begin{align}
u_1 &= \frac{1}{2} \frac{a}{\mu} \frac{\partial p_1}{\partial X} (Z-H_1)(Z-H_2) + \frac{Z-H_2}{H} \bm{v}_A \cdot \bm{e}_x, \label{u_1_solution} \\
v_1 &= \frac{1}{2} \frac{a}{\mu} \frac{\partial p_1}{\partial Y} (Z-H_1)(Z-H_2) + \frac{Z-H_2}{H} \bm{v}_A \cdot \bm{e}_y. \label{v_1_solution}
\end{align}
In addition, the corresponding Reynolds equation for the next order pressure contribution, $p_1$, is as follows:
\begin{equation}
\frac{H^3}{12} \nabla_{\perp}^2 p_1 + \frac{H^2}{4} \bigg( \frac{\lambda+1}{\lambda} \bigg) \bm{\rho} \cdot \nabla_{\perp} p_1 - \frac{\mu}{2 a} \bigg( \frac{\lambda+1}{\lambda} + \frac{H}{\rho^2} \bigg) \bm{\rho} \cdot \bm{v}_A = 0. \label{Reynolds_eqn_u1}
\end{equation}
The complete solution to Eq.~\eqref{Reynolds_eqn_u1} is of the form
\begin{equation}
p_1 (\rho,\phi) = f_{\text{p}}(\rho) + g(\rho) \cos 2\phi, \label{p1}
\end{equation}
where the particular integral is
\begin{equation}
f_{\text{p}}(\rho) = \frac{3 \mu}{4 a} \bigg( \frac{\lambda}{\lambda+1} \bigg) \frac{6H-1}{H^2} \sum_n  \bigg[ B_n W_n \big( - \bm{e} \cdot \bm{e}_z \big) \bm{e} \cdot \bm{e}_z + \frac{1}{2} B_n W_n' \big( - \bm{e} \cdot \bm{e}_z \big) ( \bm{e} \cdot \bm{e}_{x} )^2 \bigg], \label{f_p_solution}
\end{equation}
and $g(\rho)$ is the solution to a second-order differential equation (not shown). It is not necessary to solve for $g(\rho)$ since it will not contribute to the overall force on the spheres anyway. A new function, $F(\rho)$, is defined so that
\begin{equation}
F(\rho) \sum_n \bigg[ B_n W_n \big( - \bm{e} \cdot \bm{e}_z \big) \bm{e} \cdot \bm{e}_z + \frac{1}{2} B_n W_n' \big( - \bm{e} \cdot \bm{e}_z \big) ( \bm{e} \cdot \bm{e}_{x} )^2 \bigg] = \frac{a}{\mu} f_{\text{p}}(\rho),
\end{equation}
where 
\begin{equation}
F(\rho) = \frac{3}{4} \bigg( \frac{\lambda}{\lambda+1} \bigg) \frac{6H-1}{H^2}.
\end{equation}
Figure~\ref{pressure_first_and_second_order}(b) shows the dependence of $F(\rho)$ on $\rho$ for various values of $\lambda$. Importantly, these results correspond to the contribution to the pressure which is independent of $\phi$. The contribution which is proportional to $\cos 2 \phi$ will disappear upon integration with respect to $\phi$. \\

Before commencing the evaluation of the forces acting on sphere 1, it will be necessary to calculate the fluid velocity in the gap. Equations~\eqref{u_0_solution} and \eqref{v_0_solution} contain the cartesian components of the fluid velocity in the $x$ and $y$-directions, correct to first-order, with analogous solutions for the second-order case (Eqs.~\eqref{u_1_solution}-\eqref{v_1_solution}). The fluid velocity in the $\rho$ and $\phi$-directions at each order in $\epsilon$ is required, since the subsequent analysis will involve finding the rate-of-strain tensor in cylindrical coordinates. 

\subsubsection{First-order velocities}

From Eqs.~\eqref{u_0_solution}, \eqref{v_0_solution} and \eqref{leading_order_pressure}, the leading-order fluid velocity in the $\rho$ and $\phi$-directions can be written respectively as
\begin{align}
U_0 & = \bm{e} \cdot \bm{e}_{\rho} \bigg(  \frac{1}{2} \frac{a}{\mu} \frac{d q_0}{d \rho} (Z-H_1)(Z-H_2) - \frac{Z-H_2}{H} \sum_n B_n W_n \big( -\bm{e} \cdot \bm{e}_z \big) \bigg), \label{u_0_rho_dir} \\
V_0 &= \bm{e} \cdot \bm{e}_{\phi} \bigg(  \frac{1}{2} \frac{a}{\mu} \frac{q_0}{\rho} (Z-H_1)(Z-H_2) - \frac{Z-H_2}{H} \sum_n B_n W_n \big( -\bm{e} \cdot \bm{e}_z \big) \bigg). \label{u_0_phi_dir}
\end{align}
The components $u_{0,\rho}$, $u_{0,\phi}$ and $u_{0,z}$  are defined so that $u_{0,\rho}(\rho,z) \bm{e} \cdot \bm{e}_{\rho}=U_0$, $u_{0,\phi}(\rho,z) \bm{e} \cdot \bm{e}_{\phi}=V_0$ and $u_{0,z}(\rho,z) \bm{e} \cdot \bm{e}_{\rho}=\epsilon^{1/2} W_0$. The new function $v_M$ is defined so that
\begin{equation}
v_M \sum_n B_n W_n \big( -\bm{e} \cdot \bm{e}_z \big) = u_M, \quad \forall \ \text{subscripts} \ M. \label{definition_Q_0_and_v_M}
\end{equation}
The fluid velocities can thus be expressed in the following way:
\begin{equation}
U_0 = \bm{e} \cdot \bm{e}_{\rho} \ v_{0,\rho} \sum_n B_n W_n \big( -\bm{e} \cdot \bm{e}_z \big), \qquad V_0 = \bm{e} \cdot \bm{e}_{\phi} \ v_{0,\phi} \sum_n B_n W_n \big( -\bm{e} \cdot \bm{e}_z \big).
\end{equation}

\subsubsection{Second-order velocities}

The second-order pressure distribution is given by $p_1(\rho,\phi)=f_{\text{p}}(\rho) + g(\rho) \cos 2 \phi$, where $f_{\text{p}}(\rho)$ and $g(\rho)$ are known functions. The second-order fluid velocity in the $\rho$ and $\phi$-directions can be written as
\begin{align}
U_1 &=  \frac{1}{2} \frac{a}{\mu} (Z-H_1)(Z-H_2) \bigg( \frac{d f}{d \rho} + \frac{d g}{d \rho} \cos 2\phi \bigg) + \frac{Z-H_2}{H} \bm{e}_{\rho} \cdot \bm{v}_A \label{u_1_rho_dir2}, \\
V_1 &=  - \frac{a}{\mu} (Z-H_1)(Z-H_2) \frac{g}{\rho} \sin 2 \phi + \frac{Z-H_2}{H} \bm{e}_{\phi} \cdot \bm{v}_A.\label{u_1_phi_dir2}
\end{align}

\subsubsection{Tangential force}

The component of the force acting tangential to the surface of sphere 1 in the gap region is given by $d F_x = \bm{e}_x \cdot \big[ \bm{\sigma} \cdot \bm{n} \big] dA$, where $\bm{\sigma} = -p \textbf{I} + 2 \mu \boldsymbol \varepsilon$ is the stress tensor and the particle surface is defined as $A$. It can be seen that
\begin{align}
\bm{e}_x \cdot \big[ \bm{\sigma} \cdot \bm{n} \big] &= -p \sin \theta \cos \phi + 2 \mu \Big( \sin \theta (\varepsilon_{\rho^* \rho^*} \bm{e}_{\rho} + \varepsilon_{\phi \rho^*} \bm{e}_{\phi}) + \cos \theta (\varepsilon_{\rho^* z} \bm{e}_{\rho} + \varepsilon_{\phi z} \bm{e}_{\phi}) \Big) \cdot \bm{e}_x
\end{align}
and so
\begin{align}
dF_x &= \bigg[ - \big(\epsilon^{-3/2} p_0 +  \epsilon^{-1} p_1 +  \epsilon^{-1/2} p_2 + \mathcal{O}(1)\big) \sin \theta \cos \phi  \nonumber \\
& + 2 \mu \Big( \sin \theta \cos \phi \ \varepsilon_{\rho^* \rho^*} - \sin \theta \sin \phi \ \varepsilon_{\phi \rho^*}  + \cos \theta \cos \phi \ \varepsilon_{\rho^* z} - \cos \theta \sin \phi \ \varepsilon_{\phi z} \Big) \bigg] dA. \label{dF_x element}
\end{align}
In the region between the spheres, $\pi - \theta \ll 1$. Since $\sin \theta = \rho^*/a = \epsilon^{1/2} \rho$, it is appropriate to change variables according to $\rho = \epsilon^{-1/2} \sin \theta$, and so $d \rho = \epsilon^{-1/2} \sqrt{1- \epsilon \rho^2} d \theta$. In addition, $\cos \theta = -1 + \rho^2 \epsilon/2 + \mathcal{O}(\epsilon^2)$. It follows that
\begin{equation}
dA = a^2 \sin \theta \ d \phi \ d \theta = \frac{\rho a^2 \epsilon}{\sqrt{1-\epsilon \rho^2}} \ d\phi \ d\rho.
\end{equation}
The required components of the rate-of-strain tensor, $\bm{\varepsilon}$, can be evaluated using the expressions for the fluid velocities. The force element in the $x$-direction is found to be
\begin{equation}
dF_x = \bigg[ -\rho Q_0 \cos^2 \phi - \frac{\mu}{a} \bigg(\cos^2 \phi \frac{\partial v_{0,\rho}}{\partial Z} + \sin^2 \phi \frac{\partial v_{0,\phi}}{\partial Z} \bigg) \bigg] \rho a^2 \bm{e} \cdot \bm{e}_x \sum_n B_n W_n \big( -\bm{e} \cdot \bm{e}_z \big) \ d\phi \ d\rho
\end{equation}
and so
\begin{equation}
F_x = \pi a^2 \bm{e} \cdot \bm{e}_x \sum_n B_n W_n \big( -\bm{e} \cdot \bm{e}_z \big) \int_{0}^{\rho_0} \bigg[ -Q_0 \rho^2 - \frac{\mu \rho}{a} \bigg(  \frac{\partial v_{0,\rho}}{\partial Z} +  \frac{\partial v_{0,\phi}}{\partial Z} \bigg) \bigg] \ d \rho + \mathcal{O}(\epsilon^{1/2}).
\end{equation}
Since the form of $v_{0,\rho}$ and $v_{0,\phi}$ are known, the leading-order force can be calculated:
\begin{equation}
F_x = \pi a^2 \bm{e} \cdot \bm{e}_x \sum_n B_n W_n \big( -\bm{e} \cdot \bm{e}_z \big) \int_{0}^{\rho_0} \bigg[ -Q_0 \rho^2 - \frac{H}{2} \big(Q_0 \rho \big)' +  \frac{2 \mu}{a} \frac{\rho}{H}  \bigg] \ d \rho, \label{F_x sphere1 integral}
\end{equation}
where $\rho_0$ denotes the extent of the lubrication region. The three integrals can be evaluated analytically. The inner solution must now be matched with the outer solution in order to find a suitable value for the boundary of the lubrication region, $\rho_0$. Using $\rho_0 = D/\epsilon^{1/2}$ where $D \sim \mathcal{O}(1)$, it follows that $\log \rho_0^2 \sim - \log \epsilon$. With this in mind, the tangential force can be determined:
\begin{equation}
F_x^{(1)} = -\frac{4}{5} \mu \pi a \ \bm{e} \cdot \bm{e}_x \frac{\lambda (\lambda+4)}{(\lambda+1)^2} \sum_n B_n W_n \big( -\bm{e} \cdot \bm{e}_z \big)  \big( \log \epsilon + \mathcal{O}(1) \big). \label{F_x sphere1}
\end{equation}
This expression in Eq.~\eqref{F_x sphere1} represents a small correction to the results presented in ref \cite{Ishikawa2006}. However, the results for identically sized squirmers ($\lambda=1$) -- the situation of most practical interest -- are unchanged. The process of calculating the tangential force on sphere 2 is almost identical to the former case, with only a few small modifications to the analysis being necessary. Since $p_0 = p_0 (\rho,\phi)$ is independent of $Z$, the pressure at the surface of sphere 2 is the same as at the surface of sphere 1. Analysis reveals that the tangential force exerted on sphere 2 is $F_x^{(2)} = - F_x^{(1)}$.

\subsubsection{Normal force}

Consider the normal force component acting on sphere 1, given by
\begin{equation}
dF_z = \bm{e}_z \cdot \big[ \bm{\sigma} \cdot \bm{n} \big] dA = \bigg[ -p \cos \theta + 2 \mu \big( \varepsilon_{zz} \cos \theta + \varepsilon_{z \rho^*} \sin \theta \big) \bigg] dA. \label{dF_z element}
\end{equation}
Although the above equation has a dependence on $\theta$, the formulation will continue in cylindrical coordinates. The force element can be written as
\begin{align}
dF_z &= \Bigg[ -\big(\epsilon^{-3/2} p_0 +  \epsilon^{-1} p_1 +  \epsilon^{-1/2} p_2 + \mathcal{O}(1)\big) \cos \theta + \frac{2 \mu}{a} \bigg[ \bigg(\epsilon^{-1/2} \frac{\partial w_0}{\partial Z} + \frac{\partial w_1}{\partial Z} + \mathcal{O}(\epsilon^{1/2})\bigg) \cos \theta \nonumber \\
& + \frac{1}{2} \bigg( \bigg( \epsilon^{-1} \frac{\partial U_0}{\partial Z} + \epsilon^{-1/2} \frac{\partial U_1}{\partial Z} + \mathcal{O}(1)\bigg) +  \bigg( \frac{\partial w_0}{\partial \rho} + \epsilon^{1/2} \frac{\partial w_1}{\partial \rho} + \mathcal{O}(\epsilon) \bigg) \bigg) \sin \theta \bigg] \Bigg] dA, \label{dF_z_expanded}
\end{align}
where
\begin{align}
U &= U_0 + \epsilon^{1/2} U_1 + \epsilon U_2 + \mathcal{O}(\epsilon^{3/2}) \\
U_k &= u_k \cos \phi + v_k \sin \phi, \quad \text{for} \quad k=0,1,2,\ldots
\end{align}
The change of variables defined by $\rho = \epsilon^{-1/2} \sin \theta$ is again utilized. Since $\cos \theta = -1 + \rho^2 \epsilon/2 + \mathcal{O}(\epsilon^2)$, the dependence of Eq.~\eqref{dF_z_expanded} on $\theta$ can be removed, and it is found that the contributions to the force at various orders in $\epsilon$ are given by
\begin{align}
& \mathcal{O}(\epsilon^{-1/2}): \qquad F_z = a^2 \int_{0}^{\rho_0} \int_{0}^{2\pi} \rho \ p_0 \ d \phi d \rho, \label{dF_z_component1} \\
& \mathcal{O}(1): \hspace{0.1in} \quad \qquad F_z = a^2 \int_{0}^{\rho_0} \int_{0}^{2\pi} \rho \ p_1 \ d \phi d \rho, \label{dF_z_component2} \\
& \mathcal{O}(\epsilon^{1/2}):  \quad \qquad F_z =  a^2 \int_{0}^{\rho_0} \int_{0}^{2\pi} \bigg(\rho \ p_2 + \frac{\rho \mu}{a} \Big(\rho \frac{\partial U_0}{\partial Z} - 2 \frac{\partial w_0}{\partial Z} \Big) \bigg) \ d \phi d \rho, \label{dF_z_component3} \\
& \mathcal{O}(\epsilon):  \quad \quad \qquad  F_z = a^2 \int_{0}^{\rho_0} \int_{0}^{2\pi} \bigg(\rho \ p_3 + \frac{\rho \mu}{a} \Big(\rho \frac{\partial U_1}{\partial Z} - 2 \frac{\partial w_1}{\partial Z} \Big) \bigg) \ d \phi d \rho. \label{dF_z_component4} 
\end{align}
Since $p_0$ is proportional to $\cos \phi$, the integral in Eq.~\eqref{dF_z_component1} is identically zero. Consider now the next-order contribution to the normal force, given in Eq.~\eqref{dF_z_component2}. The function $p_1$ has already been found and is given in Eq.~\eqref{p1}. The term in Eq.~\eqref{p1} which is proportional to $\cos 2 \phi$ will be zero upon integration with respect to $\phi$. Thus, Eq.~\eqref{dF_z_component2} can be expressed solely in terms of $f_{\text{p}}(\rho)$:
\begin{equation}
F_z = 2 \pi a^2 \int_{0}^{\rho_0} \rho \ f_{\text{p}}(\rho) \  d \rho, \label{F_z integral}
\end{equation}
where $f_{\text{p}}(\rho)$ is defined in Eq.~\eqref{f_p_solution}. 
The corresponding integral can be evaluated analytically and subsequently expanded asymptotically for $\rho_0 \gg 1$. Performing the same matching procedure as in the preceding analysis for the tangential forces, where $\rho_0 = D/\epsilon^{1/2}$, it is found that the total force exerted on sphere 1 in the $z$-direction is
\begin{equation}
F_z^{(1)} = -9 \mu \pi a \frac{\lambda^2}{(\lambda+1)^2} \sum_n \bigg[ B_n W_n \big( - \bm{e} \cdot \bm{e}_z \big) \bm{e} \cdot \bm{e}_z + \frac{1}{2} B_n W_n' \big( - \bm{e} \cdot \bm{e}_z \big) ( \bm{e} \cdot \bm{e}_{x} )^2 \bigg] \big( \log \epsilon + \mathcal{O}(1) \big). \label{F_z sphere1}
\end{equation}
The force exerted on sphere 2 in the $z$-direction can also be found, and is given by $F_z^{(2)} = - F_z^{(1)}$. The normal force contribution was overlooked in ref \cite{Ishikawa2006}, but will form an important component of our subsequent analysis.

\subsubsection{Torque}

The torque element exerted on sphere 1 in the $y$-direction is given by
\begin{equation}
d T_y = \bm{n} \cdot \bm{e}_z \ dF_x - \bm{n} \cdot \bm{e}_x \ dF_z = a \cos \theta \ dF_x - a \sin \theta \cos \phi \ dF_z.
\end{equation}
Noting that $\varepsilon_{\rho^* z} = \varepsilon_{z \rho^*}$ and remembering that $\rho = \epsilon^{-1/2} \sin \theta$ on sphere 1, the leading-order contribution to $T_y$ can be found to be
\begin{equation}
T_y = \mu \pi a^2 \ \bm{e} \cdot \bm{e}_x \sum_n B_n W_n \big( -\bm{e} \cdot \bm{e}_z \big) \int_{0}^{\rho_0}  \rho \bigg[ \frac{\partial v_{0,\rho}}{\partial Z} + \frac{\partial v_{0,\phi}}{\partial Z} \bigg] d\rho,
\end{equation}
which can be evaluated and simplified to yield
\begin{equation}
T_y^{(1)} = \frac{16 \lambda}{5 (\lambda+1)} \mu \pi a^2 \ \bm{e} \cdot \bm{e}_x \sum_n B_n W_n \big( -\bm{e} \cdot \bm{e}_z \big) \big( \log \epsilon + \mathcal{O}(1) \big). \label{T_y sphere1}
\end{equation}
By symmetry, the torque $T_x$ is precisely equal to zero. The torque in the $z$-direction can be evaluated, however it is found that $T_z = \mathcal{O}(\epsilon)$ so this need not be pursued. The torque exerted on sphere 2 in the $y$-direction is readily computable using the results of the preceding results. Most of the working is the same as before, and it is found that
\begin{equation}
T_y^{(2)} = \frac{4 \lambda^2}{5 (\lambda+1)} \mu \pi a^2 \ \bm{e} \cdot \bm{e}_x \sum_n B_n W_n \big( -\bm{e} \cdot \bm{e}_z \big) \big( \log \epsilon + \mathcal{O}(1) \big). \label{T_y sphere2}
\end{equation}
There exists an extra factor of $\lambda$ compared to the results for sphere 1, arising from the discrepancy between their radii. It is also worth noting that for $\lambda = 1$, the torque exerted on sphere 2 is one quarter times that exerted on sphere 1. Normal gradients in the fluid velocity are greater at the surface of the squirmer than they are at the boundary of the no-slip sphere, giving rise to this somewhat counterintuitive result.

\section{Monolayer of squirmers in an unbounded fluid: unrestricted motion} \label{SI_outofplane}

To this point, the position and orientation vectors of the squirmers have been restricted to lie in the $x$-$z$ plane. Importantly, it was found that for certain combinations of parameters, there exists a critical value of $G_{bh}$ above which small perturbations to the orientations and positions of the squirmers will decay, and the equilibrium monolayer configuration is restored. In particular, this stability depended on the presence of the repulsive force. Without it, the monolayer was unstable, even for large $G_{bh}$. Perturbations to the position and orientation vectors will now be permitted to be out of the plane of the equilibrium configuration depicted in Fig.~\ref{diamond_domain}. For this the full matrix-vector equation presented in Eq.~\eqref{matrix-vector_equation3} is constructed and solved for the linear and angular velocities. The coordinates and orientation of each squirmer can be represented by $\bm{r}_i = (x_i,y_i,z_i)$ and $\bm{e}_i = (\sin \theta \cos \phi, \sin \theta \sin \phi, \cos \theta)$ respectively. The time-stepping is done in the same fashion as before, using the Runge-Kutta-Fehlberg method (RKF45).

Firstly, consider the effects of perturbing the positions of the squirmers in any of the three spatial dimensions. For these simulations, the initial orientations of the squirmers were left unperturbed, pointing in the $z$-direction. The positions were given a perturbation in a random direction, of magnitude $\epsilon_0 a/100$. The results corresponding to $G_{bh}=50$ and $\beta=1$ are presented, since this parameter combination previously gave rise to a stable monolayer.

\begin{figure}[htp]
\begin{center}
\includegraphics[width=0.85\textwidth]{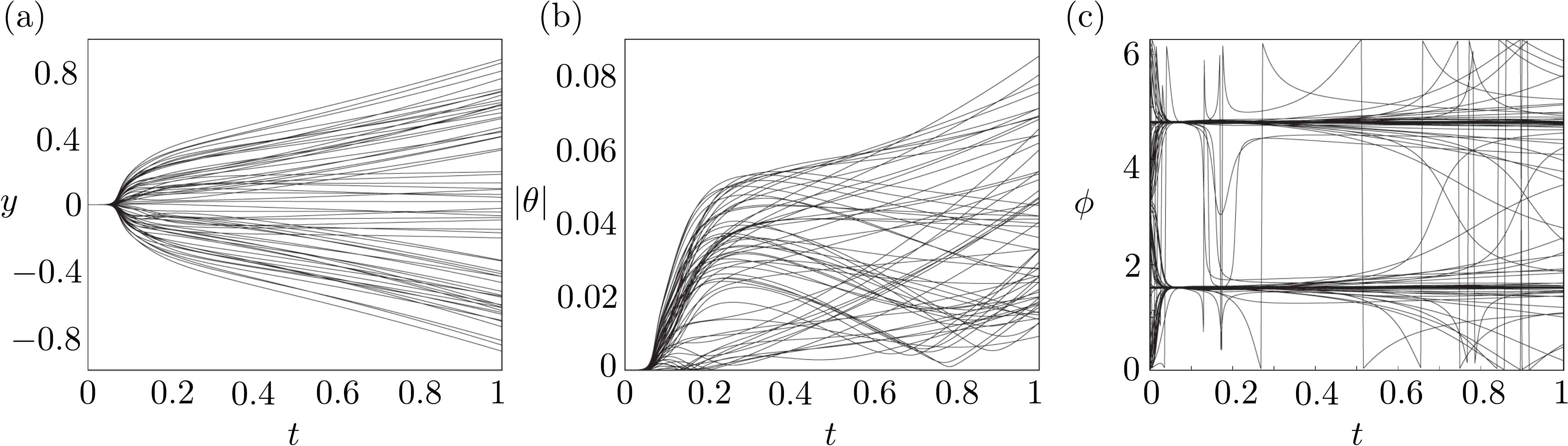}
      \caption{Figure showing the $y$-coordinate and orientation ($\theta$ and $\phi$) of every squirmer in the $8 \times 8$ diamond, as functions of time. For this simulation, every squirmer was initially given a random perturbation to its position, with amplitude $\epsilon_0 a /100$. Results have been computed with $G_{bh}=50$, $\epsilon_0=2 \times 10^{-3}$, $\kappa_1=1$, $\kappa_2=10^3$ and $\beta=1$ over the interval $t \in [0,1]$.}
  \label{outofplanestability}
  \end{center}
\end{figure}

It is clear from Fig.~\ref{outofplanestability}(a) that the small perturbations to the squirmers' positions quickly grow, with the monolayer beginning to drift apart in the $y$-direction. At larger times, the squirmers move with constant velocity in the $y$-direction. Although the initial perturbation is only in the squirmers' positions, Fig.~\ref{outofplanestability}(b) illustrates that their orientations quickly deviate from their equilibrium value. Results have also been computed for the case where only the orientations of the squirmers in the equilibrium monolayer are perturbed. This gives rise to a set of results which are qualitatively the same as those presented in Fig.~\ref{outofplanestability}. When motion of the squirmers was limited to lie in the plane of the monolayer, the configuration was stable (see green curves in Fig.~\ref{zeta_vs_time_rep}(b)). It is clear, however, that the extra degrees of freedom in the squirmers' position and orientation serve to destabilize this monolayer. This result is in accordance with previous findings \cite{Ishikawa2008a} where elaborate coherent structures formed in 2D were not observed in full 3D simulations.

\section{Monolayer of squirmers between rigid walls} \label{SI_between_two_walls}

It has already been shown that the uniform monolayer of squirmers in an unbounded fluid is unstable when subjected to small translational or rotational perturbations in the direction perpendicular to their plane (SI Section~\ref{SI_outofplane}). This out-of-plane motion is now suppressed by including the two walls. To achieve this, Equation~\eqref{all_forces_and_torques} is modified by the inclusion of additional forces and torques given by:
\begin{align}
\left( \begin{array}{c}
\bar{\bm{F}}_1 \\
\vdots \\
\bar{\bm{F}}_n \\
\hline
\bar{\bm{T}}_1 \\
\vdots \\
\bar{\bm{T}}_n \\
\end{array} \right)^{\text{wall}}
&=
\left( \begin{array}{c|c}
\textbf{M}_1^{\text{wall}} & \textbf{M}_2^{\text{wall}} \\
\hline
\textbf{M}_3^{\text{wall}} & \textbf{M}_4^{\text{wall}} \\
\end{array} \right)
\left( \begin{array}{c}
\bm{V}_1 \\
\vdots \\
\bm{V}_n \\
\hline
a \bm{\omega}_1 \\
\vdots \\
a \bm{\omega}_n \\
\end{array} \right)
+
\left( \begin{array}{c}
\bar{\bm{F}}_1 \\
\vdots \\
\bar{\bm{F}}_n \\
\hline
\bar{\bm{T}}_1 \\
\vdots \\
\bar{\bm{T}}_n \\
\end{array} \right)^{{}^{\text{sq}}_{\text{wall}}} 
+
\left( \begin{array}{c}
\bar{\bm{F}}_1 \\
\vdots \\
\bar{\bm{F}}_n \\
\hline
0 \\
\vdots \\
0 \\
\end{array} \right)^{{}^{\text{rep}}_{\text{wall}}}. \label{all_forces_and_torques_wall}
\end{align}
The matrices $\textbf{M}_i^{\text{wall}}$ are block-diagonal since the forces and torques acting on any sphere are independent of the linear and angular velocities of any other sphere. By again demanding that the squirmers are all force- and torque-free, it is possible to construct a matrix-vector equation of the same form as Eq.~\eqref{matrix-vector_equation3}. In the absence of the two walls, a reference frame had to be chosen with some arbitrary velocity (see Eq.~\eqref{net_velocity_zero}). Upon inclusion of the walls, the choice is no longer arbitrary, since the forces and torques arising through interaction with the planes depend on their velocities relative to the squirmers. 

\begin{figure}[htp!]
\begin{center}
\includegraphics[width=\textwidth]{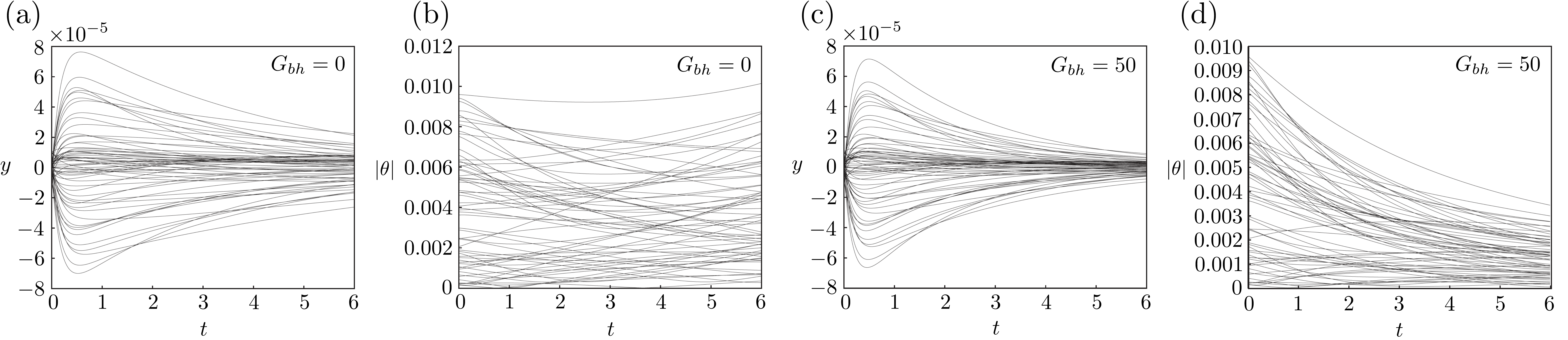}
\caption{Figure showing the $y$-coordinate and orientation from vertical, $\theta$, of every squirmer in the $8 \times 8$ diamond, as a function of time. For these simulations, every squirmer was initially given a random perturbation to its position and orientation, with amplitudes $\epsilon_0 a /100$ and $1/100$ respectively. (a) and (b) have been computed with $G_{bh}=0$ while (c) and (d) correspond to $G_{bh}=50$. Results have been computed with $\epsilon_0=2 \times 10^{-3}$, $\epsilon_0^{\text{wall}}=5 \times 10^{-3}$, $\kappa_1 = \kappa_1^{\text{wall}} = 1$, $\kappa_2 = \kappa_2^{\text{wall}}=10^3$ and $\beta=1$ over the interval $t \in [0,6]$.}
\label{3D_wall_Gbh_0and50}
\end{center}
\end{figure}

Consider the consequences of initiating [small] perturbations to the positions and orientations of the squirmers in the monolayer between the walls. The results corresponding to $\beta = 1$, $\epsilon_0 = 2 \times 10^{-3}$ and $\epsilon_0^{\text{wall}} = 5 \times 10^{-3}$ are presented, and the differences between the cases involving $G_{bh}=0$ and $G_{bh}=50$ examined. From Figs.~\ref{3D_wall_Gbh_0and50}(a) and \ref{3D_wall_Gbh_0and50}(c) it is clear that the walls have the effect of stabilizing the positions of the squirmers. In this respect, they help to maintain the structure of the monolayer. This phenomenon is insensitive to the value of $G_{bh}$ used. Figures~\ref{3D_wall_Gbh_0and50}(b) and \ref{3D_wall_Gbh_0and50}(d) demonstrate the effect that changing $G_{bh}$ has on the system. For $G_{bh}=0$, the orientation of the squirmers grows with time while for $G_{bh} = 50$ the squirmers in the monolayer are ultimately restored to their equilibrium orientation. This is the same behavior that was exhibited in Fig.~\ref{zeta_vs_time_rep} where a critical value of $G_{bh}$ was required to maintain the equilibrium configuration of the monolayer. The difference, now, is that the nearby plane walls hold the monolayer together in the $y$-direction instead of having to ignore motion in that direction.

\end{document}